\documentclass[12pt]{article}

\newif\iffigs\figstrue

\usepackage{epsfig,latexsym}
\usepackage{amsmath}
\usepackage{color}
\usepackage{verbatim}
\usepackage{mathrsfs}
\usepackage{amssymb}

\DeclareMathAlphabet{\mathpzc}{OT1}{pzc}{m}{it}

 \csname
@addtoreset\endcsname{equation}{section}

\def\gz0{\gamma^{0}}

\def\scs#1{\section{\sc #1}}

\def\beq{\begin{equation}}
\newcommand{\eeq}[1]{\label{#1}\end{equation}}
\def\bea{\begin{eqnarray}}
\newcommand{\eea}[1]{\label{#1}\end{eqnarray}}
\def\ba{\begin{array}}
\def\ea{\end{array}}
\def\bec{\begin{center}}
\def\ec{\end{center}}
\def\ba{\begin{align}}
\def\ena{\end{align}}

\def\12{\frac{1}{2}}

\usepackage{slashed}

\usepackage{float}

\usepackage{booktabs}
\usepackage{caption}

\usepackage[mathscr]{euscript}

\usepackage{color}
\usepackage{graphicx}
\usepackage{epsf}
\usepackage{graphicx,epsfig}
\usepackage{amsmath}

\usepackage{multirow}

\usepackage{amsmath, amsthm, amssymb}

\usepackage{epsfig}
\usepackage{cite}
\usepackage{color,colordvi}

\newcounter{hran}

\makeatletter
\renewcommand\section{\@startsection {section}{1}{\z@}%
                               {-3.5ex \@plus -1ex \@minus -.2ex}%
                               {2.3ex \@plus.2ex}%
                               {\normalfont\large\bfseries}}
\makeatother

\vspace{0.5cm}

\newcommand{\bi}{\begin{itemize}}
\newcommand{\ei}{\end{itemize}}

\begin{document}
\thispagestyle{empty}
\begin{flushright}
\end{flushright}


\begin{center}


{\Large\sc Low--$\ell$ CMB from String--Scale SUSY Breaking?}\\


\vspace{60pt}
{\sc A.~Sagnotti}\\[15pt]

{\sl\small Scuola Normale Superiore and INFN\\
Piazza dei Cavalieri \ 7\\ 56126 Pisa \ ITALY  \\ }
\vspace{6pt}

\vspace{70pt}

 {\sc\large Abstract} \end{center}

\baselineskip=14pt

\noindent Models of inflation are instructive playgrounds for supersymmetry breaking in Supergravity and String Theory. In particular, combinations of branes and orientifolds that are not mutually BPS can lead to \emph{brane supersymmetry breaking}, a phenomenon where non--linear realizations are accompanied, in tachyon--free vacua, by the emergence of steep exponential potentials. When combined with milder terms, these exponentials can lead to slow--roll after a fast ascent and a turning point. This leaves behind distinctive patterns of scalar perturbations, where pre--inflationary peaks can lie well apart from an almost scale invariant profile. I review recent attempts to connect these power spectra to the low--$\ell$ CMB, and a corresponding one--parameter extension of $\Lambda$CDM with a low--frequency cut $\Delta$. A detailed likelihood analysis led to $\Delta = (0.351 \pm 0.114) \times 10^{-3} \, \mbox{Mpc}^{-1}$, at $99.4\%$ confidence level, in an extended Galactic mask with $f_{sky}=39\%$, to be compared with a nearby value at $88.5\%$ in the standard {\sc Planck} 2015 mask with $f_{sky}=94\%$. In these scenarios one would be confronted, in the CMB, with relics of an epoch of deceleration that preceded the onset of slow--roll.


\vspace{50pt}
\noindent {\sl Based on the presentations at the International School for Subnuclear Physics, 53rd Course,
``The Future of our Physics Including New Frontiers'',
Erice, June 24 -- July 3 2015, and at ``Physics on the Riviera 2015'', Sestri Levante, September 16--18 2015}

\vfill

\noindent

\setcounter{page}{1}

\pagebreak

\newpage
\scs{Introduction}

It is often stated that Physics thrives on crises, and Cosmology \cite{cosmology} is somehow coming to the rescue these days, with a wealth of data which add new motivations to explore the possible impact, in the Early Universe, of ideas that emerged in Particle Physics and are still awaiting an experimental vindication.

The study of CMB anisotropies is a notable example in this respect. It has already contributed significantly to granting Cosmology an unprecedented accuracy that rests, to a large extent, on the underlying theory of cosmological perturbations. Rather involved to begin with, this theory has the virtue of resting largely on linear phenomena and brings to the forefront, within the theory of inflation \cite{inflation}, measurable aspects of the merging of Quantum Mechanics and General Relativity. The highlight of all this is a slight tilt of the CMB primordial power spectrum \cite{cm},
\beq
{\cal P}_R(k) \ \sim \ k^{n_s-1} \ ,
\eeq{1}
for which {\sc Planck} recently obtained the result \cite{planck15}
\beq
n_s \ = \ 0.968\ \pm \ 0.006 \ ,
\eeq{2}
so that an additional, concrete window opens up on inflation, one of the most enticing ideas that have emerged in Cosmology. This early phase of accelerated expansion, proposed as a natural way to account for the flatness of the Universe and for the apparent paucity of its inhomogeneities, could have been realized if a scalar field, the inflaton, underwent a slow--roll motion in the very early Universe, thus simulating the repulsive effect of a positive cosmological constant before releasing its energy into the vacuum.

Once a slow--roll phase finds its place in Cosmology, it becomes natural to inquire how it started. This type of investigation suffered, over the years, from the lack of compelling scenarios to address it that can be blamed, to some extent at least, on our incomplete grasp of String Theory \cite{stringtheory}. At any rate, as a departure from de Sitter space, inflation brings along naturally high--scale supersymmetry breaking in scenarios suggested by String Theory and Supergravity \cite{supergravity}.  The models that will guide the ensuing discussion were inspired by \emph{brane supersymmetry breaking} \cite{bsb}, a peculiar mechanism that shows up in String Theory, in orientifold constructions \cite{orientifolds} involving collections of branes and orientifolds that are not mutually BPS.  In its simplest manifestation, the ten--dimensional Sugimoto model of \cite{bsb}, the introduction of $O9_+$ orientifold planes requires the presence of anti D9-branes, and this induces the breaking of supersymmetry at the string scale and the emergence of non--linear realizations \cite{dm} in vacua that are classically stable, in that they are free of tachyon excitations. In the following I shall concentrate on some amusing scenarios that these constructions suggest for the Early Universe, trying to draw from them some lessons for the low--$\ell$ CMB. There is a lot to learn, however, from systems of this type in many respects, since unbalanced forces exist among their extended objects \footnote{See \cite{polchinski_15} for a recent review with stimulating critical assessments.}, and non--BPS systems of BPS branes seem to provide an ideal entry point into the dynamics of String Theory with broken supersymmetry.

We made an early encounter with this problem in the early 1990's, with M.~Bianchi and G.~Pradisi, which is reported in \cite{erice92}. We had efficient world--sheet rules to deal with supersymmetric systems, but occasionally they seemed to run into trouble and to require what, in the current geometric language of \cite{polchinski}, would amount to negative numbers of branes. $O_+$-planes had already made an indirect appearance in the toroidal compactifications of \cite{orientifolds}, and when their nature was elucidated in \cite{witten_oplus} these types of systems were reconsidered in connection with the issue of supersymmetry breaking. The $O_+$-planes are BPS objects, characterized by equal values for their tension $T>0$ and charge $Q>0$, and it is the need to cancel the latter that brings about anti D-branes, anti-BPS objects with $T>0$ and $Q<0$. The unbalanced tension reflects itself, in the string frame, in the emergence of an exponential potential for the dilaton $\phi$,
\beq
V \ \sim T \, e^{\,-\,\phi} \ ,
\eeq{3}
whose exponent is determined by the Polyakov expansion.
\begin{figure}[ht]
\centering
\includegraphics[width=70mm]{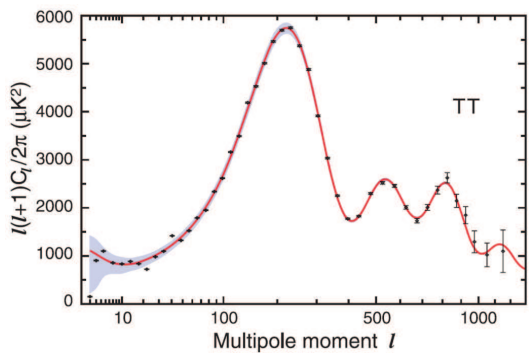}
\caption{\small Deviations from the concordance model $\Lambda$CDM show up in the low--$\ell$ end of this nice WMAP9 plot of the CMB angular power spectrum. Most of them are foreshadowed by Cosmic variance, but for a low quadrupole and a dip around $\ell=20$.}
\label{fig:CMB_tot}
\end{figure}

Since the potential \eqref{3} has no critical point, resummation procedures are to be invoked to connect these systems to flat backgrounds \cite{dnps}, but Cosmology and inflation provide a more natural setting. Mild exponential potentials supporting inflation were indeed studied since the 1980's \cite{lm,exponential}, but remarkably the exponent in eq.~\eqref{13} sits precisely at a (steep) \emph{critical value} where it starts to become impossible, for $\phi$, to emerge from an initial singularity while descending the potential \cite{dks} (at least in the two--derivative approximation \cite{cd}).
Hence, the \emph{climbing scalar} generically reaches up to a point where it reverts its motion and starts to descend. If a critical exponential is accompanied by other contributions capable of supporting an inflationary phase, a picture builds up for the onset of inflation with a burst of pre--inflationary slow-roll around the turning point. This can introduce, in the power spectrum of scalar perturbations, a pre--inflationary peak well apart from the almost scalar invariant profile \eqref{1} that is eventually attained \cite{ks}, in contrast with what happens if the scalar attains slow-roll directly, in which case the peak is contiguous to the almost scale invariant profile. The study of a decelerating inflaton in connection with quadrupole depression, in different types of scenarios, attracted some attention during the last decade or so, and an (incomplete) list of references can be found in \cite{standard_peak,standard_peak_recent}.

\begin{figure}[ht]
\centering
\begin{tabular}{ccc}
\includegraphics[width=30mm]{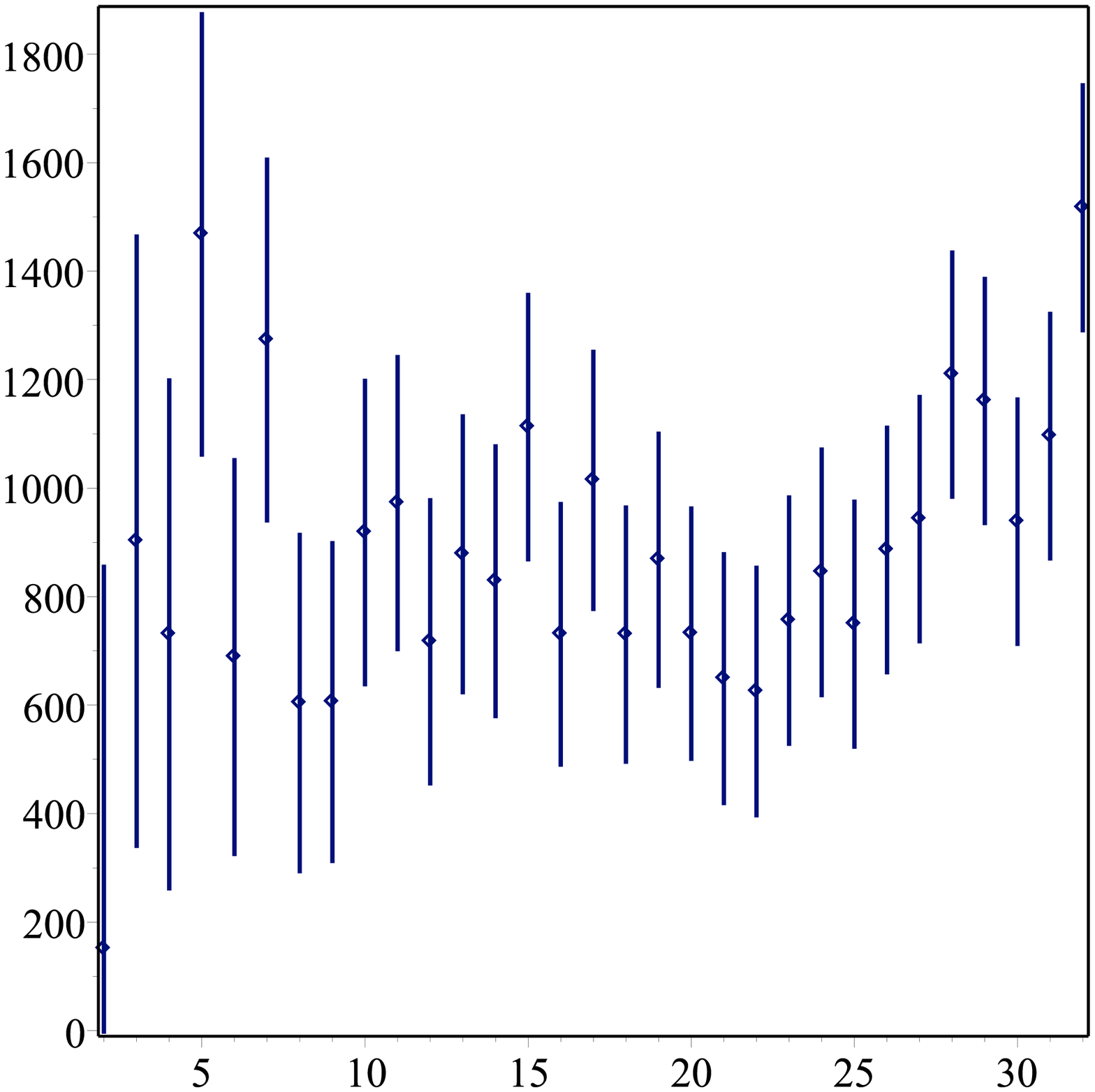} \qquad\quad &
\includegraphics[width=30mm]{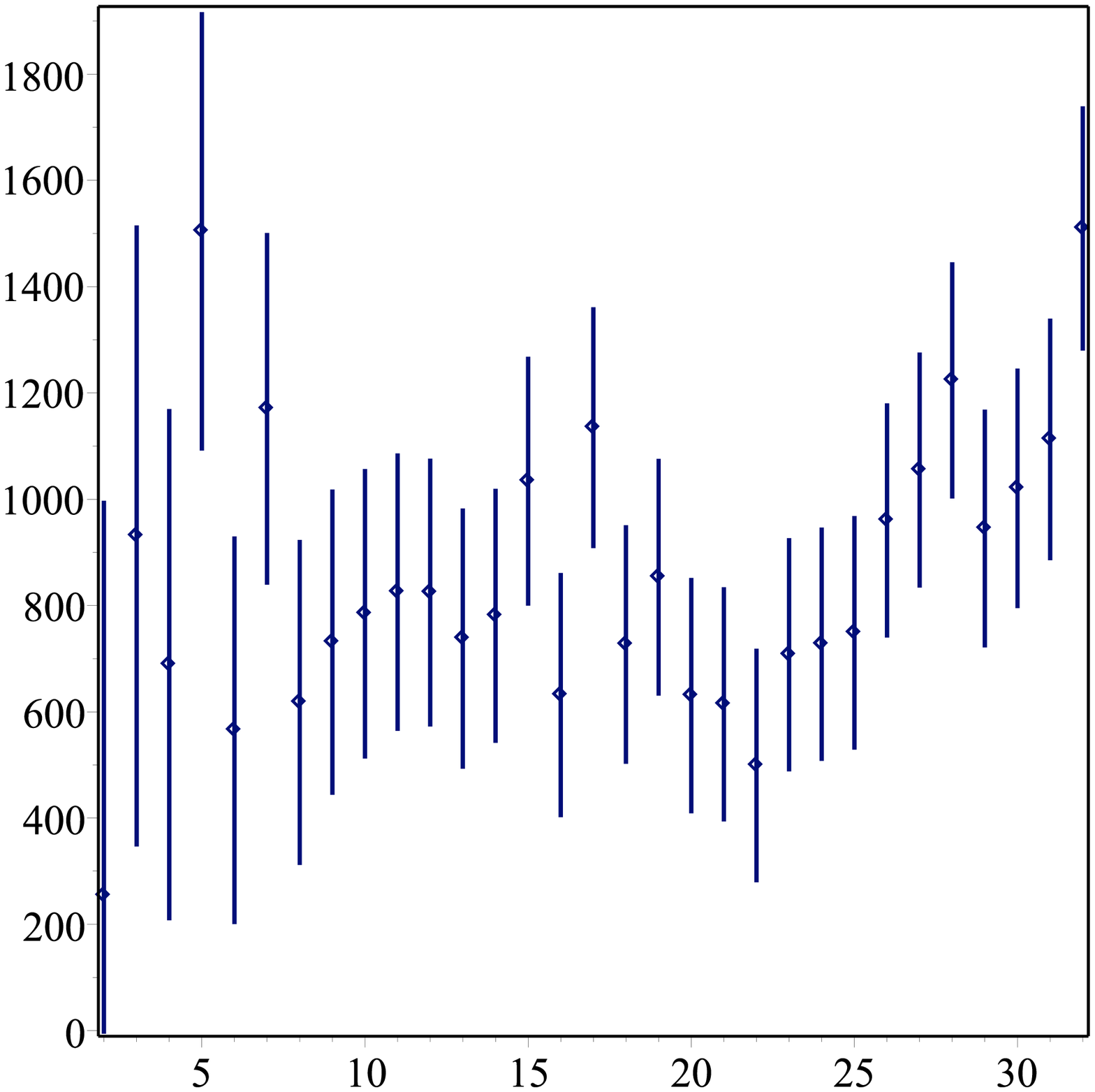} \qquad\quad &
\includegraphics[width=30mm]{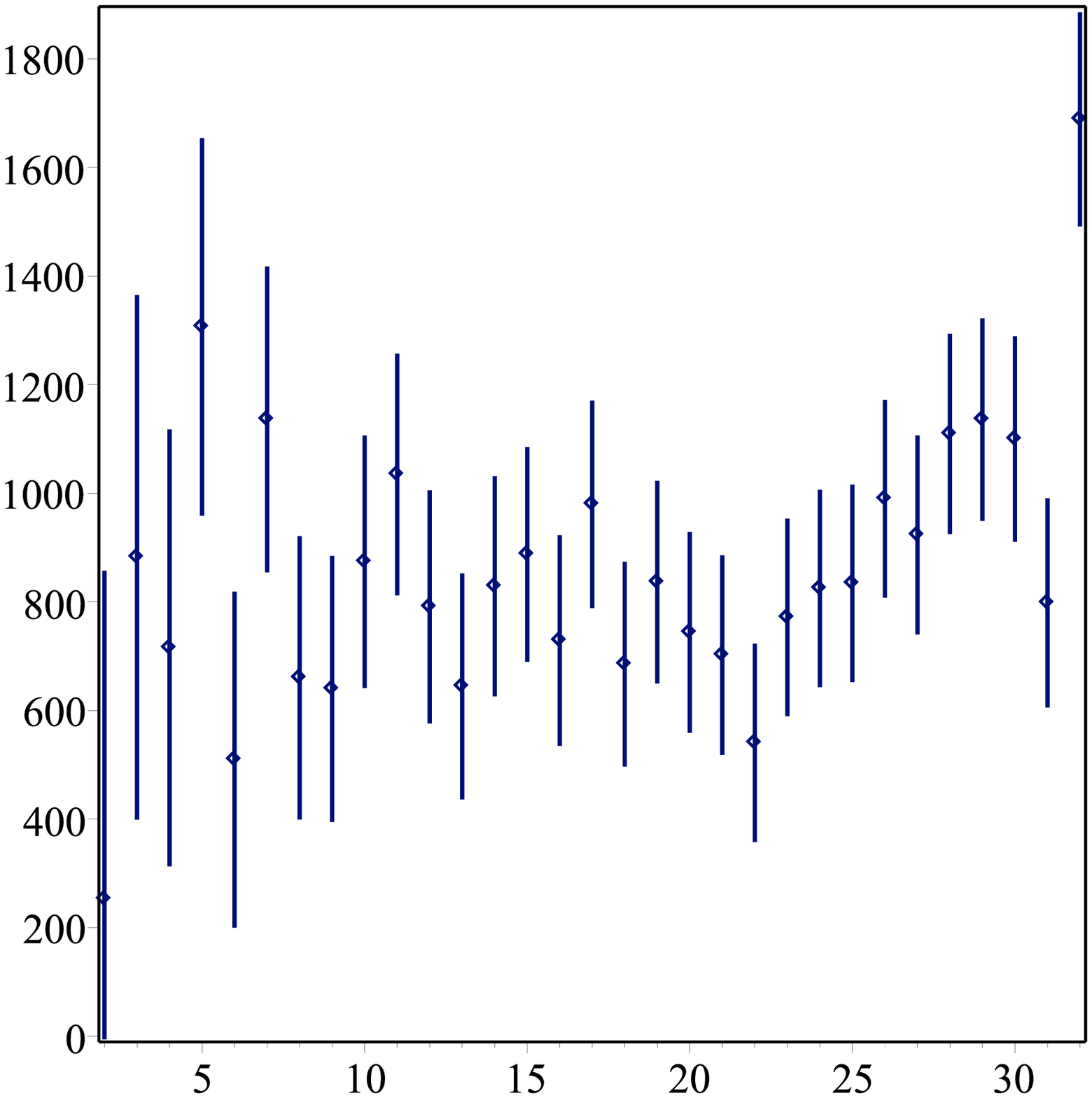}\vspace{12pt}
\end{tabular}
\caption{\small A closer look at the low--$\ell$ end ($\ell \leq 32$) according to WMAP9 (left), {\sc Planck} 2013 (center) and {\sc Planck} 2015 (right).}
\label{fig:compared_CMB}
\end{figure}

The key issue that I shall address here is whether the types of feature emerging from brane supersymmetry breaking, or at least signs of a decelerating inflaton, could be hiding behind the anomalies that have long emerged in the low--$\ell$ tail of the observed CMB angular power spectrum (APS). To this end, let me begin with fig.~\ref{fig:CMB_tot}, which is relatively simple to read since it displays, in the region of interest, only a few binned data without error bars. Leaving aside momentarily the shadowed region, the reader will spot a very low quadrupole, followed by a large oscillation around $\ell=5$, a milder peak and a sizable dip around $\ell=20$. A closer look at the first multipoles displayed in fig.~\ref{fig:compared_CMB} reveals that matters are more murky, but a low quadrupole, a sizable peak following it around $\ell=5$, a smaller one around $\ell=15$ and a dip around $\ell=20$ appear common to all these results.

Two facts make this type of analysis fascinating. To begin with, the theory of inflation associates the low--$\ell$ range to the earliest epochs that are accessible via the CMB. Moreover, the lower end of the CMB anisotropy spectrum contains, in principle, pristine information about primordial power spectra of scalar perturbations. The Universe behaves somehow as an optical filter for these frequencies, so that \cite{cmb_slow}
\beq
A_\ell \ \sim \ \ell(\ell+1) \ \int \frac{dk}{k} \ P_R(k) \ j_\ell(k \Delta \eta)^2 \ \simeq \ P_R\left(k=\frac{\ell}{\Delta \eta}\right) \ ,
\eeq{4}
where the last step is justified since the spherical Bessel functions build up a narrow kernel. There is, however, a big complication: Cosmic variance, an overall incertitude on low--$\ell$ data that reflects the limited information that we can collect on them, is the reason for the shadowed region in the left end of fig.~\ref{fig:CMB_tot}. We can indeed detect a single realization of the anisotropy pattern from our special vantage point in the Universe while, for any $\ell$, the information at our disposal rests on $2\ell+1$ values of the ``magnetic quantum number'' $m \in [-\ell,\ell]$, with expected fluctuations that behave proportionally to $\frac{1}{\sqrt{2\ell+1}}$. Caution is more than legitimate, although in scenarios that originate from higher dimensions, which admittedly I am unable to advocate at present in a convincing form, one could be dealing with many more data. For instance, 55 independent data would be contributing to the quadrupole in $D=10$! Keeping cosmic variance well in mind, I shall thus continue to examine the low--$\ell$ anomalies, drawing some inspiration from String Theory and Supergravity.

Let me stress, however, that the preceding considerations, and in particular the models of high--scale supersymmetry breaking, do not force pre--inflationary dynamics to affect the low--$\ell$ end of the CMB anisotropy. Rather, as the recent onset of an accelerated phase, the possibility of accessing some information related to the onset of inflation would represent a sort of cosmic coincidence. This type of scenario is in principle compatible with a short inflation, and if it were correct some information on how inflation started would clearly make the whole picture more concrete. Future experiments might grant further support to these ideas, or perhaps rule them out altogether. An important datum, in this respect, is that in scenarios of this type the tensor--to--scalar ratio $r$ tends to exceed, at low frequencies, the value that it attains in the slow--roll phase \cite{dkps}.

In order to minimize the overlap with some preceding reviews, I shall begin in Section
\ref{sec:climbing} with a quick discussion of the climbing phenomenon, illustrating it also with reference to a simple analytic model introduced in \cite{fss}. In Section \ref{sec:chi2} I shall describe some simple inflationary models combining these features and corresponding $\chi^{2}$ tests related to first 30 multipoles. The highlight will be that it is simple, and even natural, to capture a low quadrupole, a large peak around $\ell=5$ and even a dip around $\ell=20$, via a climbing scalar. However, all these models share the unpleasing feature of yielding power spectra where the eventual almost scale invariant profile is approached too slowly. In  Section \ref{sec:galactic} I shall begin by reviewing some analytic Coulomb--like models of primordial power spectra that were proposed in \cite{dkps}. These well-motivated departures from eq.~\eqref{1} at low $k$ can define two--parameter extensions of $\Lambda$CDM that account for a decelerating inflaton. Their most important new parameter, a scale $\Delta$, characterizes typical energy scales at the epoch where slow--roll inflaton would have started. Its determination will bring about a crucial new ingredient, the choice of Galactic mask.

\scs{Climbing scalars, in brief} \label{sec:climbing}

Exponential potentials result in interesting cosmological solutions that have been explored since the 1980's \cite{lm,exponential}, but a key feature lies well beyond slow--roll profiles and went unnoticed for a while. This is the \emph{climbing behavior} of a minimally coupled scalar in the presence of \emph{(over)critical} exponentials \cite{dks}. In brief, a scalar field $\phi$ minimally coupled to gravity in $D$ dimensions, which evolves in a spatially flat FLRW Universe cannot emerge from an initial singularity while descending an exponential potential
\beq
V \ = \ V_0 \ e^{\,2\, \gamma\, \varphi} \ ,
\eeq{6}
if $\gamma \geq 1$. This simple description of the result applies to all $D$, albeit in terms of the non--standard field $\varphi$, which is related to a canonically normalized field $\phi$ according to
\beq
\varphi \ = \ \sqrt{\frac{D-1}{D-2}} \ \phi \ .
\eeq{7}
For example, in terms of a canonically normalized scalar field $\phi$, in four dimensions the \emph{critical} exponential potential would read $e^{\sqrt{6}\, \phi}$.

A field of this type will be called a \emph{climbing scalar}. The nature of the transition at $\gamma=1$ and the analogy with a Newtonian particle in a viscous medium were reviewed in \cite{as_13,fss}, and more recently in \cite{fs_2015_2}. Here I shall only mention, in passing, some intriguing facts, drawn largely from \cite{dks}:
\begin{itemize}
\item the Polyakov expansion implies that the ten--dimensional Sugimoto model of \cite{bsb}, the simplest realization of brane supersymmetry breaking, has precisely $\gamma=1$;

\item once the scalar is forbidden to emerge while descending the potential, the string coupling is bounded from above, rendering this behavior naturally protected against string loops (although not against $\alpha^\prime$ string corrections \cite{cd});

\item the dynamics of $\varphi$ is fully determined up to a single constant, $\varphi_0$, which gauges the impact with the steep exponential and has important effects on primordial power spectra of scalar perturbations;

\item the KKLT uplift \cite{KKLT} can be naturally ascribed to  brane supersymmetry breaking, and the climbing behavior continues to occur also in presence of a KKLT axion. Moreover, the transition at $\gamma=1$ extends to more general potentials that are dominated asymptotically by a steep exponential;

\item in lower dimensions, the dilaton mixes with the breathing mode and one of the resulting combinations, $\Phi_t$, retains a critical exponential potential for all $D<10$. If the other, $\Phi_s$, is somehow stabilized, the climbing behavior persists for $D<10$ \cite{as_13,fss};

\item if $\Phi_s$ is somehow stabilized, a space-time filling $p$-brane that couples to the dilaton, in the string frame, proportionally to $e^{\,-\,\alpha\,\phi}$, contributes in four dimensions an exponential potential  \cite{as_13,fss}
\beq
V \sim e^{\,2\,\gamma\, \varphi} \ ,
\eeq{8}
where
\beq
\gamma \ = \ \frac{1}{12} \left( p \ + \ 9 \ - \ 6\, \alpha \right) \ .
\eeq{9}
\end{itemize}

The simplest models to describe the onset of inflation via a climbing phase rest on the two--exponential potentials
\beq
{V}(\varphi) \ = \ {V}_0 \left( e^{\,2\,\varphi} \ + \ e^{\,2\,\gamma\, \varphi} \right) \ ,
\eeq{10}
which lead to a spectral index $\left(\gamma\,<\, \frac{1}{\sqrt{3}}\,\right)$~\footnote{The first model of this type was considered in \cite{dks}, and involved the non--BPS D3-brane found in \cite{dms} following \cite{sen}.}
\beq
n_s \ = \ \frac{1\ - \ 9\, \gamma^{\,2}}{1\ - \ 3\, \gamma^{\,2}} \ .
\eeq{11}
There is some entertaining numerology: one could somehow associate $\gamma=\frac{1}{12}$ to an NS-fivebrane, and this value would result in a spectral index $n_s \simeq 0.96$.

In the next section I shall compare the toy models of eq.~\eqref{10} to the low--$\ell$ end of the CMB angular power spectrum. However, in \cite{ks} we also described a more realistic modification of the Starobinsky potential of \cite{inflation},
\beq
{V}(\varphi) \ = \ {V}_0 \left[ e^{\,2\,\varphi} \ + \ \frac{1}{2} \ e^{\,2\,\gamma\, \varphi} \ + \ \left(1 \ - \ e^{\,-\,\frac{2}{3}\ \varphi} \right)^2 \right] \ ,
\eeq{12}
which can grant, with a proper choice of $\varphi_0$, about 60 $e$--folds of inflation, with a spectral index $n_s \simeq 0.96$ and a tensor to scalar ratio $r \simeq 0.15$ for a large fraction of it. The results that I shall present will be very similar in the two cases, since they rest on the behavior near the turning point, and I shall also explain how, adding to these potentials a small gaussian bump close to the steep exponential,
\beq
\Delta V \ = \ V_0 \ a_1 \, e^{\,-\,a_2(\varphi+a_3)^2} \ ,
\eeq{13}
with $a_1$ of a few percent and $a_3$ of order one, one can introduce other interesting features in power spectra, making them amusingly close to fig.~\ref{fig:compared_CMB} \cite{ks}.

I would like to conclude this section with a brief illustration of an exact solution found in \cite{fss}. It corresponds to the class of two--exponential potentials
\beq
{V}(\varphi) \ = \ {V}_0 \left( e^{\,\frac{2}{\gamma}\ \varphi} \ + \ e^{\,2\,\gamma\, \varphi} \right) \ ,
\eeq{131}
where $0<\gamma<\frac{1}{\sqrt{3}}$ in the range of interest,
which are similar to those of eq.~\eqref{10} but much steeper. Their solutions are also qualitatively similar, can be obtained analytically and exhibit clearly the origin, in this case, of the climbing phenomenon. Letting
\beq
ds^2 \, = \, e^{\,2\,{\cal A}(t)} dt^2 \, - \, e^{\,\frac{2\,{\cal A}(t)}{D-1}} \,
\mathrm{d}\mathbf{x}\cdot \mathrm{d}\mathbf{x} \ ,
\eeq{14}
which entails a peculiar choice for the time variable, a minimally coupled Einstein--scalar system leads to the reduced action principle
\beq
{\cal L} \ = \ \frac{1}{2} \left( \dot{\varphi}^{\,2} \ -
\ \dot{\cal A}^{\,2} \right) -
{\cal V}_0 \left( e^{\,2\,({\cal A}\,+\,\gamma\,\varphi)} \
+ \ e^{\,\frac{2}{\gamma}\,(\gamma\, {\cal A}\,+\,\varphi)} \right) \ ,
\eeq{15}
which can be turned into the manifestly separable one
\beq
{\cal L} \ = \ \frac{1}{2} \left( {\dot{\widehat{\varphi}}}^{\,2} \ - \ {\dot{\widehat{\cal A}}}^{\,2} \right) \ - \ {\cal V}_0 \left( e^{\,2\,\sqrt{1-\gamma^{\,2}}\, \widehat{\cal A}}\ + \ e^{\,\frac{2}{\gamma}\,\sqrt{1-\gamma^{\,2}}\, \widehat{\varphi}}\right)
\eeq{16}
via the ``Lorentz boost''
\bea
&&\widehat{\cal A}\ = \ \frac{1}{\sqrt{1-\gamma^{\,2}}} \ \Big( \cal A \ + \ \gamma \, \varphi \Big)\nonumber \ , \\
&&\widehat{\varphi}\ = \ \frac{1}{\sqrt{1-\gamma^{\,2}}} \ \Big( \varphi \ + \ \gamma \, {\cal A} \Big) \ .
\eea{17}

The key feature underlying the climbing behavior is therefore, in this case, an impenetrable barrier on which $\widehat{\varphi}$ can only impinge to undergo a reflection. The exact solution,
\beq
e^{\,\varphi} \ = \  e^{\,\varphi_0} \ \left[ \frac{\sinh(\omega \gamma \tau)}{\cosh \omega (\tau\, -\, \tau_0)}
\right]^\frac{\gamma}{1-\gamma^2} \ , \qquad e^{\,{\cal A}} \ = \ e^{\,{\cal A}_0} \ \left[ \frac{\cosh^{\gamma^{\,2}}\!\!\omega(\tau\, -\, \tau_0)}{\sinh(\omega \gamma \tau)}
\right]^\frac{1}{1-\gamma^2} \ ,
\eeq{251}
where
\beq
\omega^2 \ = \ \frac{\lambda}{\gamma}  \  \sqrt{1-\gamma^2} \ e^{\,2\, {\cal A}_0 \, \sqrt{1-\gamma^2}} \ ,
\eeq{24}
displays, as advertised, the typical hard--bounce behavior described at length in \cite{ks}.

\scs{$\chi^{\,2}$ tests of the first few multipoles ($\ell \leq 32$)} \label{sec:chi2}

Arriving at the Mukhanov--Sasaki (MS) equation was a major achievement, which took a number of years and a proper assessment of the actual observables. Many details can be found in \cite{cosmology}, but the end result that will concern us is simply
\beq
\left(\frac{d^2}{d \eta^2} \ + \ k^2 \ - \ W_S(\eta) \right) v_k(\eta) \ = \ 0 \ .
\eeq{25}
Although here one is solving an \emph{initial--value} problem, this minimal form should ring a bell to anyone familiar with one--dimensional scattering in Quantum Mechanics, especially since the final aim of the analysis is to compute, in the $\eta \to 0^-$ limit, the power spectrum
\beq
P_R(k) \ = \ \frac{k^3}{2\,\pi^2} \ \left| \frac{v_k(\eta)}{z(\eta) }\right|^2 \ .
\eeq{26}
The key difference with respect to one--dimensional scattering is transparent in the WKB limit, since the solutions are not tuned to decay after a reflecting wall. Starting, in a region where $W_s \simeq 0$, from the oscillatory Bunch--Davies solutions
\beq
v_k(\eta) \ \sim \ e^{\,-\,i\,k\,\eta} \ ,
\eeq{27}
$W_s$ builds up a Coulomb--like barrier after several $e$--folds, where the growing mode dominates, so that the familiar tunneling formula leaves way to the limiting value
\beq
v_{k}(\eta)\ \ \ \ \thicksim\!\!\!\!\!\!\!\!\!\!_{{}_{{\eta \to
0^-}}} \ \frac{1}{\sqrt[4]{
|W_S(\eta)\,-\, k^2|}} \
\exp\left(\int_{-\eta^\star}^{\eta}
\sqrt{|W_s(y)\,-\, k^2| }\, dy \right) \ .
\eeq{281}
Consequently, larger $W_S$ build up larger $P(k)$.

I can now complete this brief discussion adding some more details on the nature of $W_S$ and of the corresponding potential, $W_T$, which determines tensor perturbations. To begin with, one can prove the two limiting behaviors,
\beq
W_{S,T} \ \ \ \thicksim\!\!\!\!\!\!\!\!\!\!_{{}_{{\eta \to -
\eta_s}}}
\ - \ \frac{1}{4}\ \frac{1}{(\eta+\eta_s)^2} \ , \qquad
W_{S,T} \ \ \ \thicksim\!\!\!\!\!\!\!\!\!\!_{{}_{{\eta \to - 0^-}}}
\  \ \frac{\nu^2 - \frac{1}{4}}{\eta^2} \qquad \left( \nu \ =
\ \frac{3}{2} \ \frac{1\,-\,\gamma^2}{1\,-\,3\,\gamma^2} \right)
\eeq{28}
which manifest themselves close to the initial singularity at $\eta=-\eta_s$
and after several $e$--folds of slow--roll. These results follow from the expressions \footnote{As in \cite{ks}, our scale factor is $\exp{{\cal A}/3}$.}
\beq
W_S \ = \ \frac{1}{z} \ \frac{d^2 z}{d\,\eta^2} \ , \quad {\rm with} \quad z \ = \ e^\frac{\cal A}{3} \ \frac{d \varphi}{d \cal A}  \ ,
\eeq{30}
and
\beq
W_T \ = \ \frac{1}{\cal A} \ \frac{d^2 {\cal A}}{d\,\eta^2} \ .
\eeq{32}
\begin{figure}[ht]
\begin{center}$
\begin{array}{cccc}
\epsfig{file=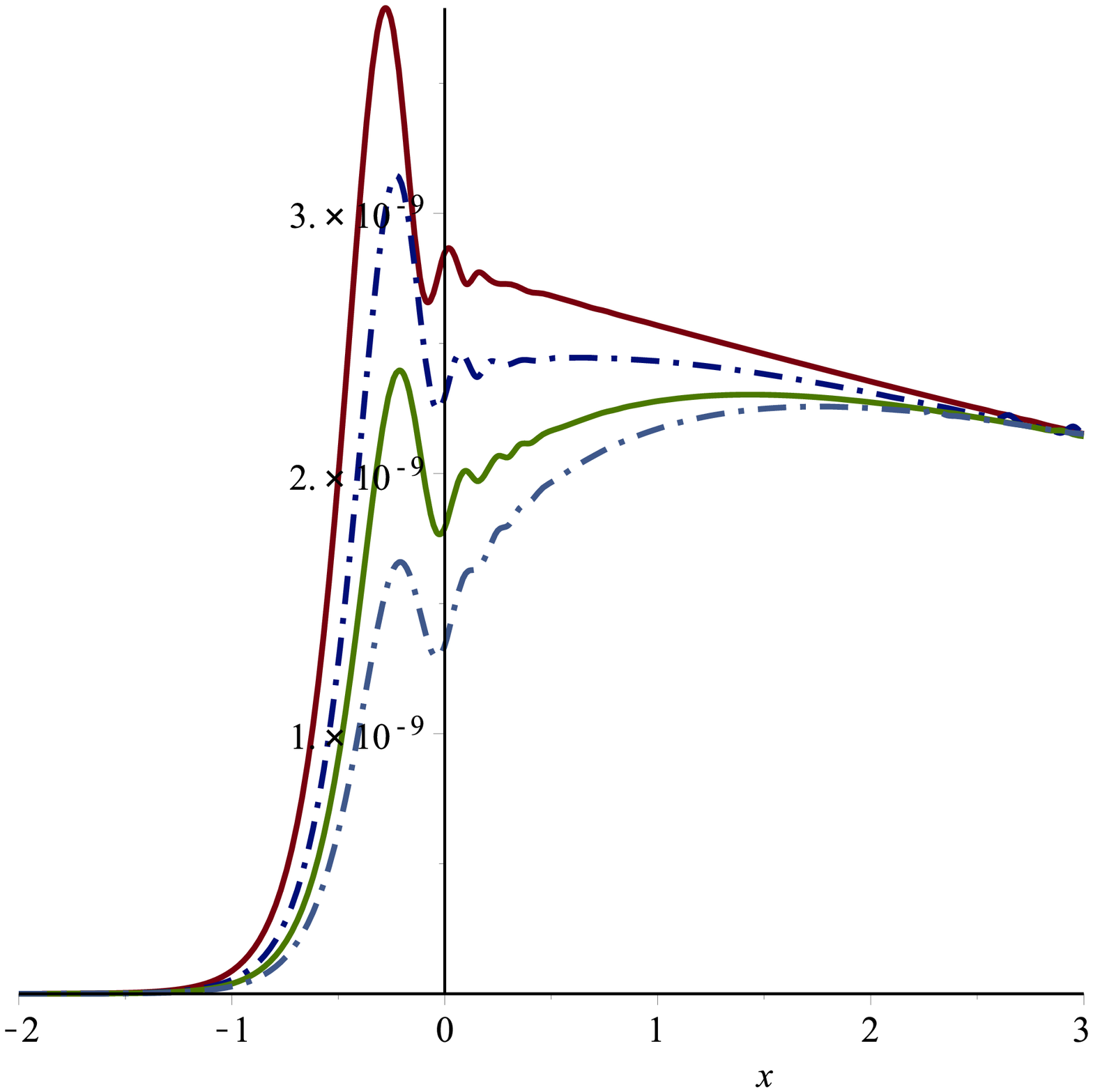, height=1in, width=1in} & \qquad
\epsfig{file=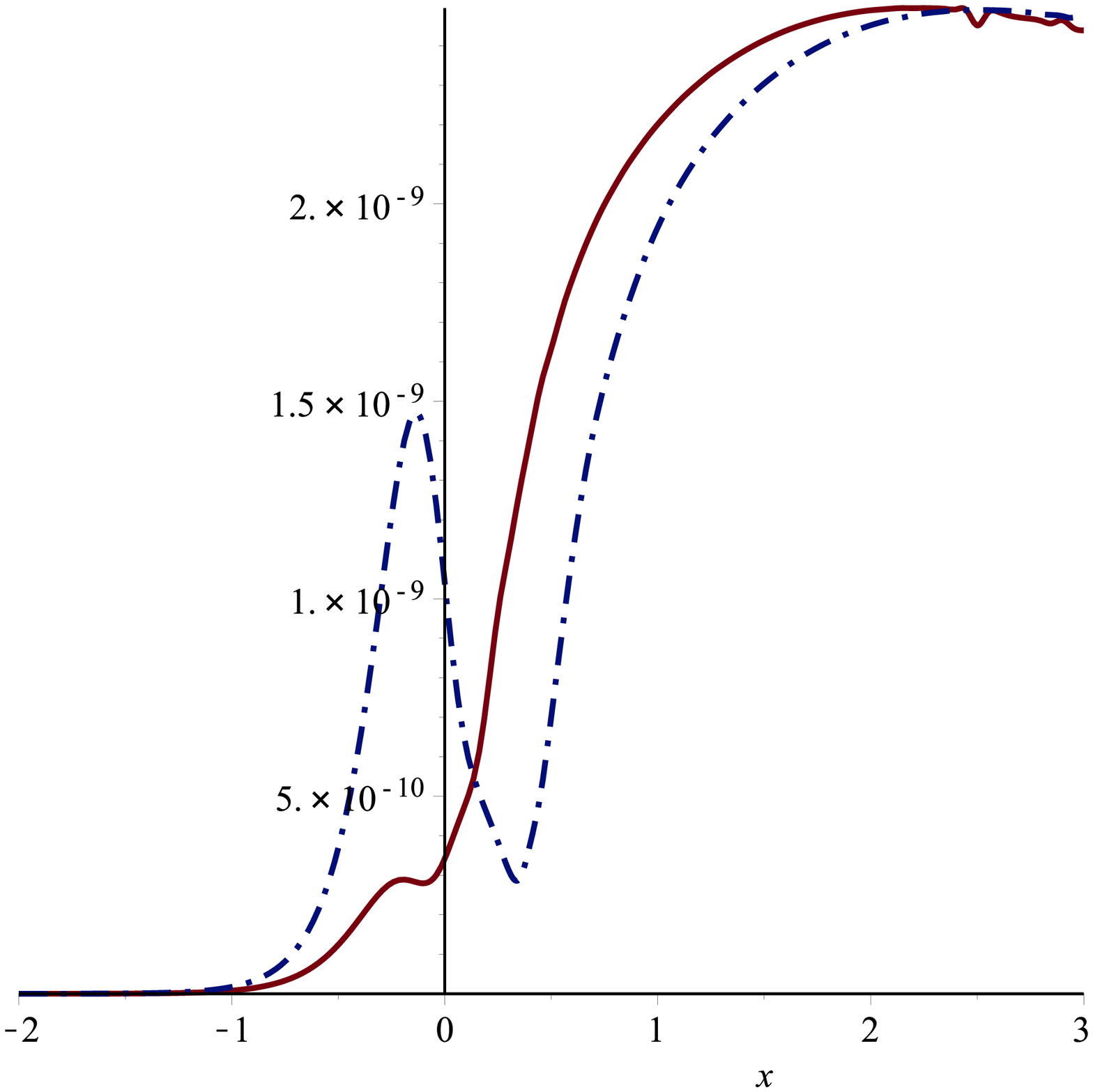, height=1in, width=1in} & \qquad
\epsfig{file=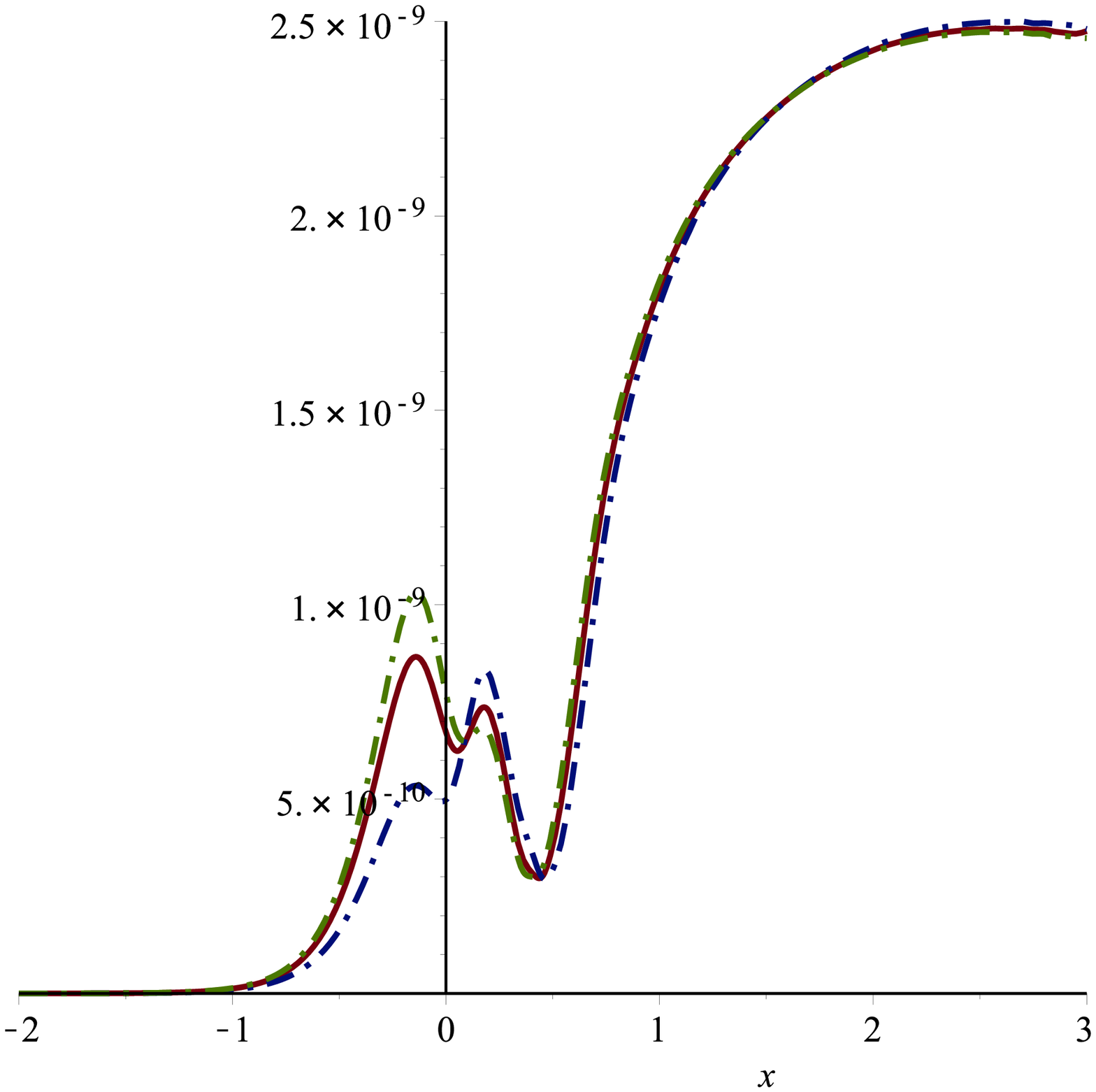, height=1in, width=1in} & \qquad \epsfig{file=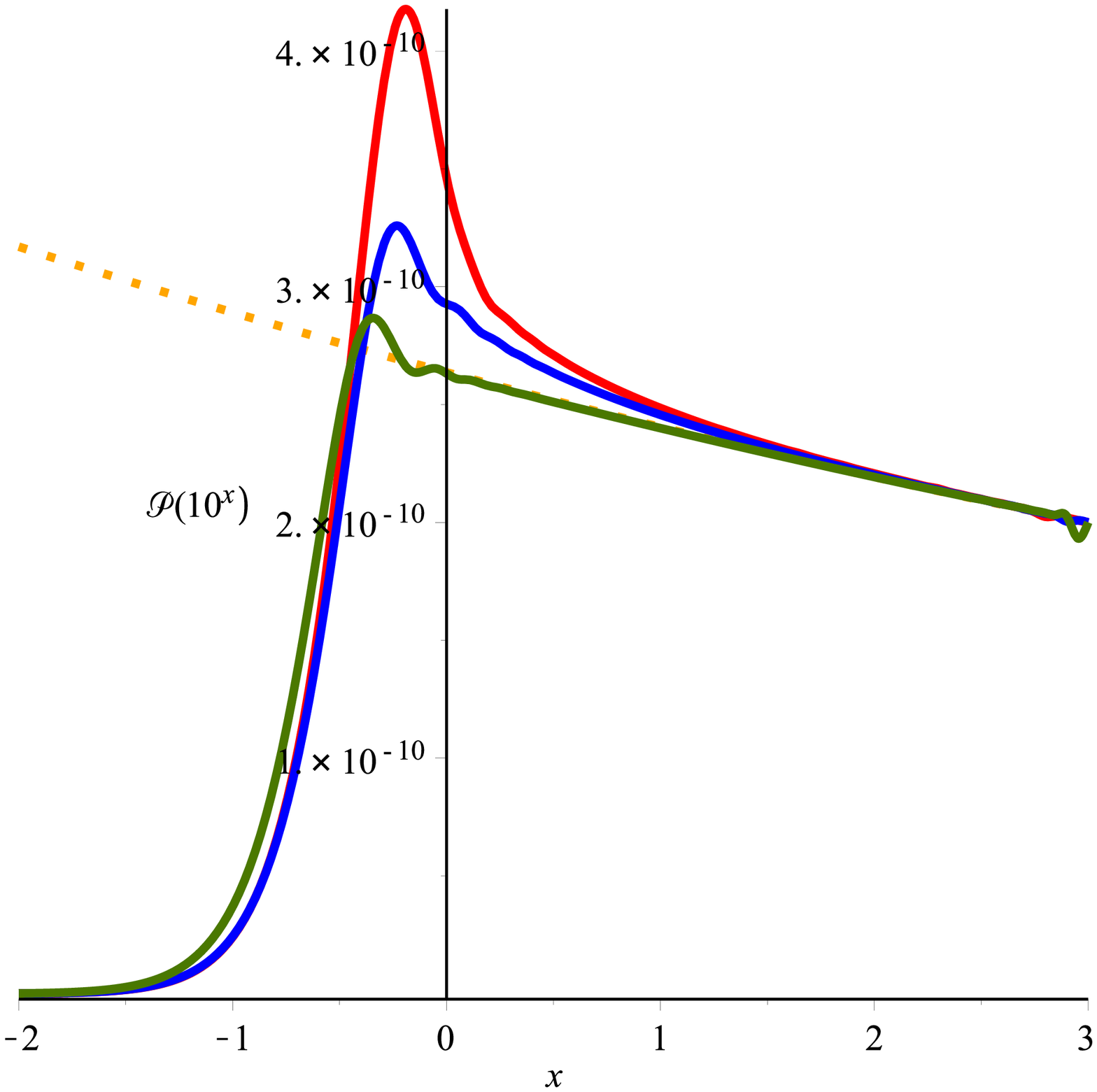, height=1in, width=1in}
\end{array}$
\end{center}
\caption{\small {\it First three:} power spectra of scalar perturbations for the double--exponential potential of eq.~\eqref{12} with a small gaussian bump as in eq.~\eqref{13}. The horizontal axis displays $x$, where $k=10^x$, and these power spectra are normalized, in an arbitrary but convenient fashion, so that they all meet as they approach the almost scale invariant profile of eq.~\eqref{1}. {\it Last:} some corresponding power spectra of tensor perturbations. }
\label{fig:double}
\end{figure}
A key lesson to be drawn from eqs.~\eqref{30} and \eqref{32} is that scalar perturbations are highly affected, in general, by the scalar dynamics, in contrast with tensor perturbations. A neat result derived in \cite{dkps}, which applies in the region where the mild exponential dominates, is that \emph{the ratio $W_S/W_T$ is less than one when the scalar is about to attain slow roll}. As a result, one can argue from the WKB expression of eq.~\eqref{281} that, in these models, the tensor--to--scalar ratio $r$ is larger, at low $k$, than its limiting value in the slow--roll region, consistently with the larger values attained by the slow--roll parameter $\epsilon$ before the scalar attains slow--roll.

I can now describe the power spectra in fig.~\ref{fig:double}, all of which approach for large $k$ the almost scale invariant profile of eq.~\eqref{1} \cite{ks}. Let me begin from the first group of plots, which obtain for values of $\varphi_0$ that are near the lower end of the range of interest. Initially, the scalar hardly meets the gaussian bump and behaves as in the presence of a single mild exponential. The power spectrum presents the typical pre--inflationary peak studied in \cite{standard_peak,standard_peak_recent}, which falls directly on the almost scale invariant profile of eq.~\eqref{1}. However, as $\varphi_0$ is increased, it begins to experience the gaussian bump, which reflects it, before it reaches its inflection, like a steep exponential would. Consequently, the pre--inflationary peak moves away from the almost scale invariant profile \cite{ks}, since the scalar slows down momentarily close to the turning point but reverts faster and faster as $\varphi_0$ is increased. Increasing $\varphi_0$ even further we thus come to the second group of plots: the peak becomes smaller and smaller, until a featureless growth is attained, along the lines of what was found in \cite{dkps} and \cite{ks} for the two--exponential case. As the scalar reaches beyond the inflection point, however, it almost comes to rest near the top and the pre--inflationary peak grows dramatically. Then, increasing $\varphi_0$ further, the scalar reaches beyond the top, so that it can finally approach the steep exponential and undergo a reflection nearby. Now it slows down twice, in climbing the gaussian bump and then near the steep exponential, and as a result \emph{two pre--inflationary peaks} build up, followed by a \emph{furrow} which reflects the fast exit from the gaussian bump. This last type of feature is well represented in the third group of plots. As we shall see shortly, it appears favoured by the low--$\ell$ CMB, where the furrow gives rise to a dip around $\ell=20$. Finally, some typical spectra of tensor perturbations are collected in the last entry of fig.~\ref{fig:double}. They undergo less dramatic changes as $\varphi_0$ is varied, as I had anticipated.
\begin{figure}[ht]
\begin{center}$
\begin{array}{ccc}
\epsfig{file=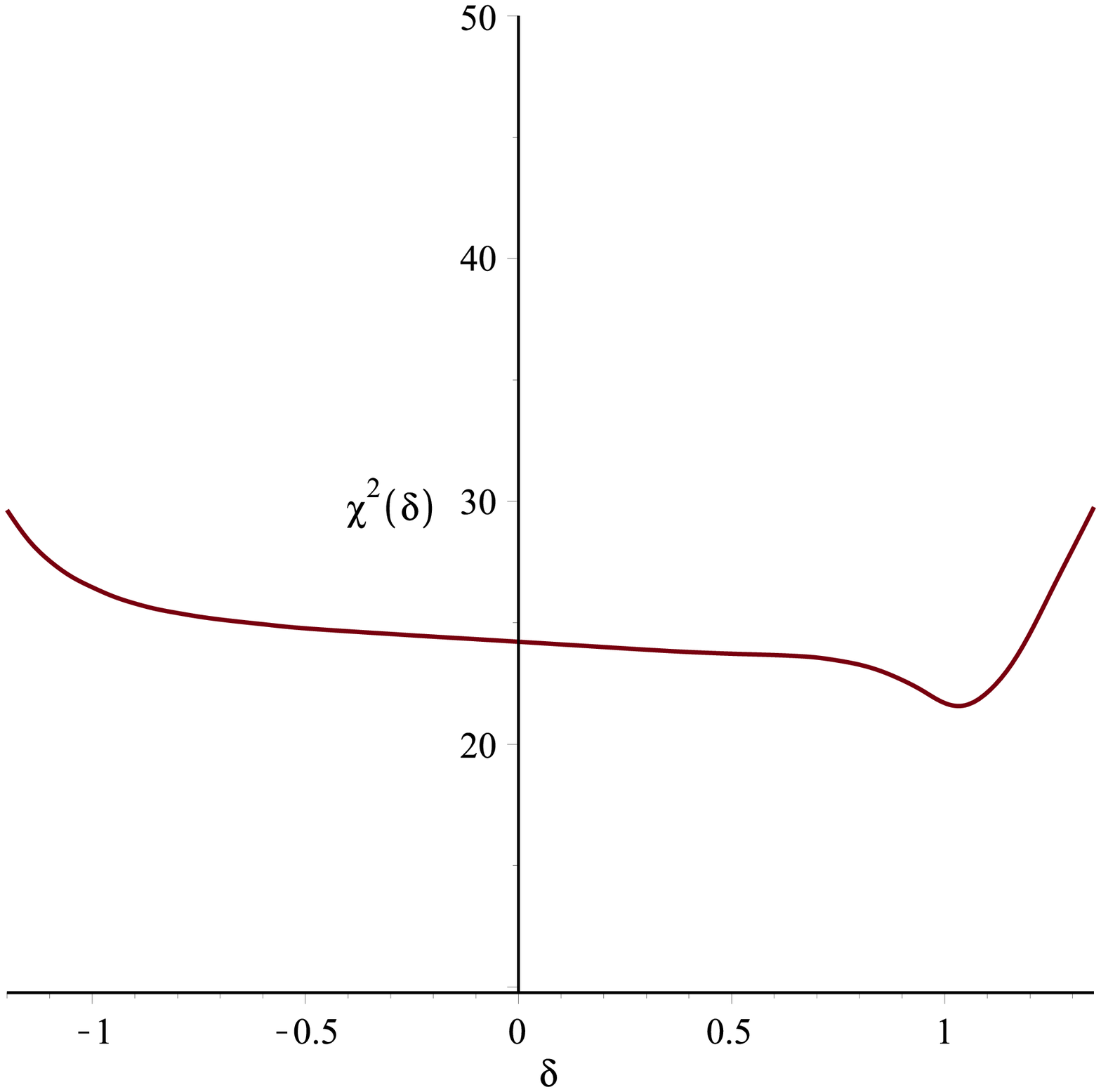, height=1.1in, width=1.1in}& \qquad\quad
\epsfig{file=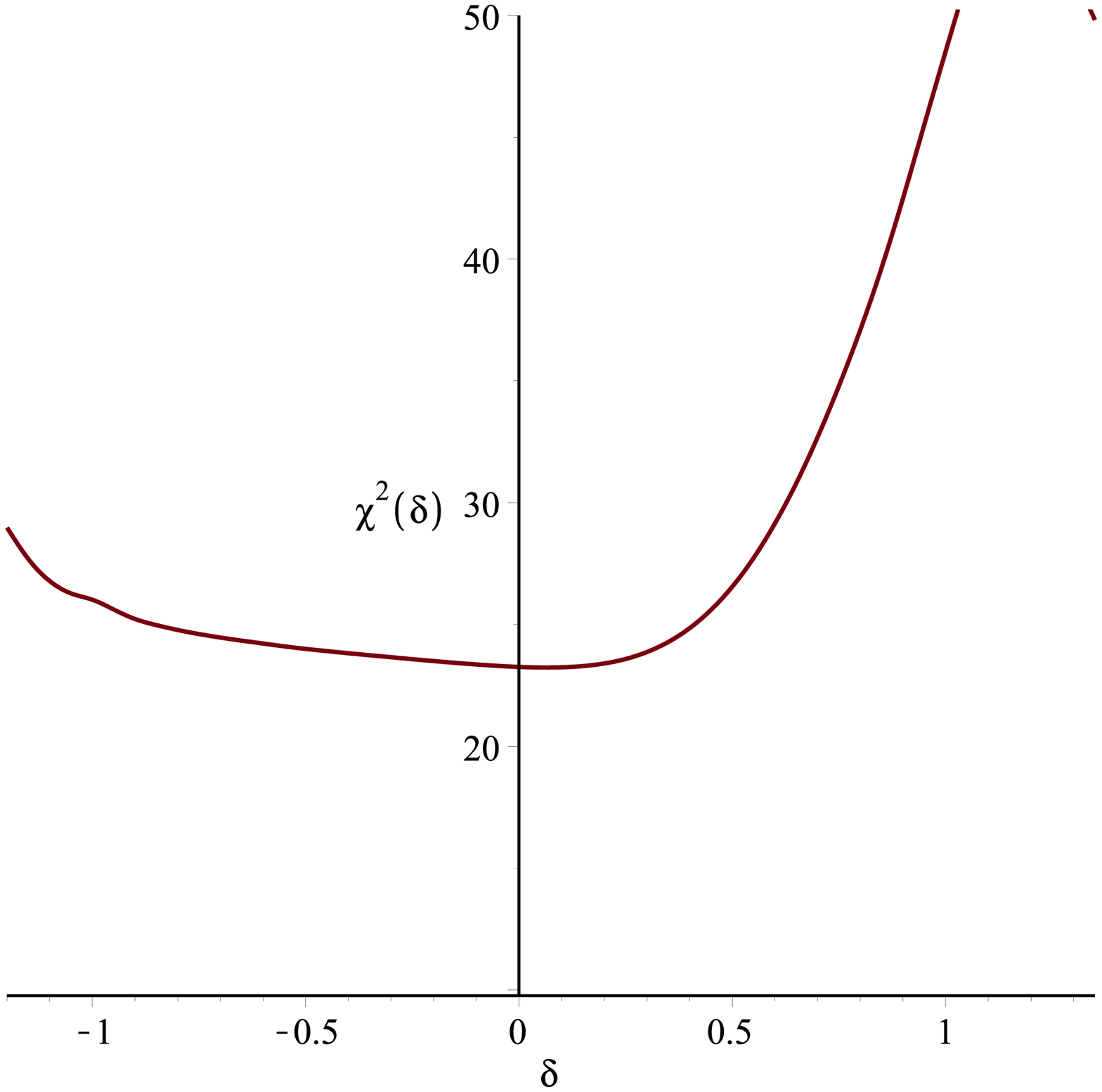, height=1.1in, width=1.1in}& \qquad\quad
\epsfig{file=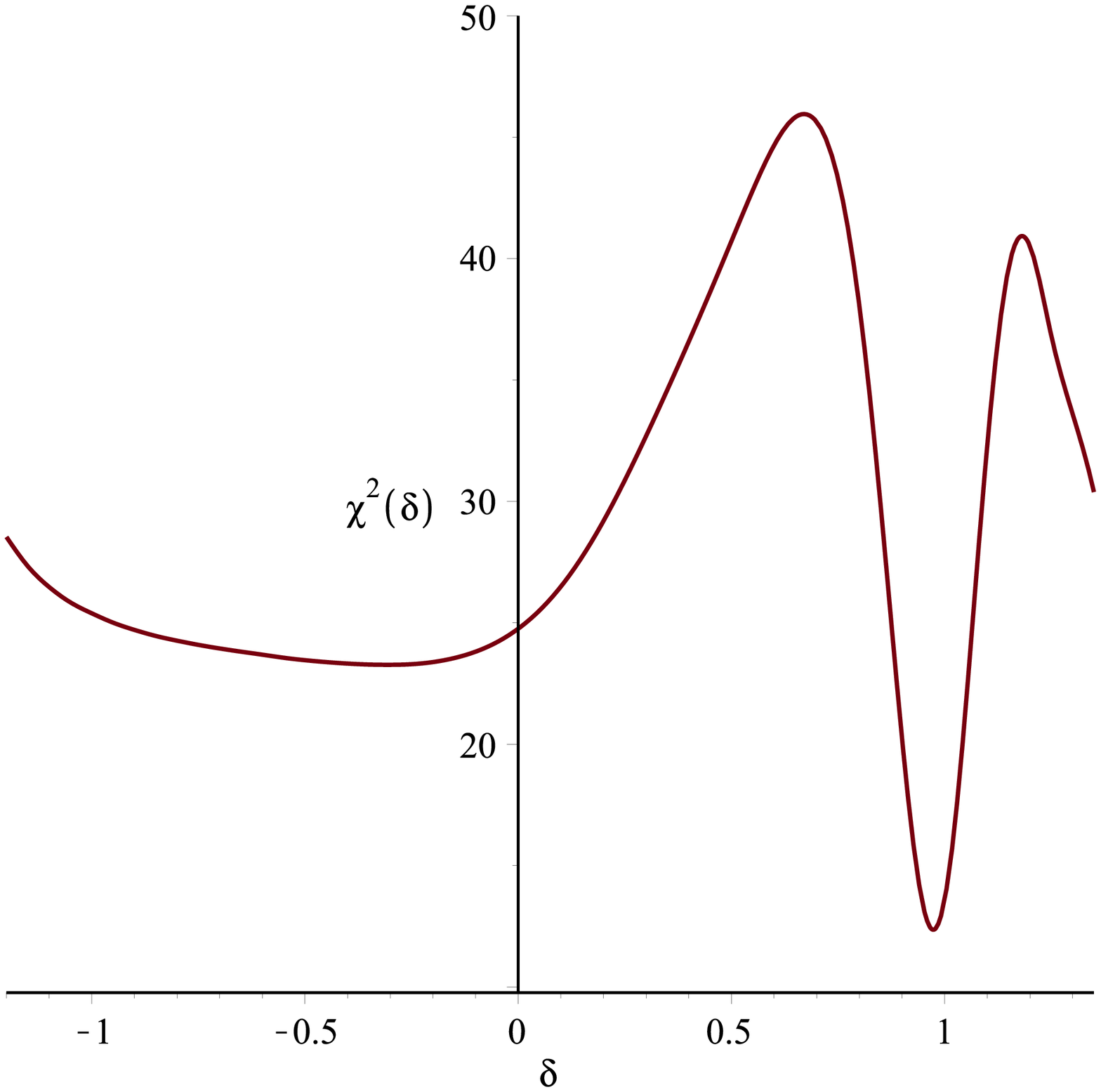, height=1.1in, width=1.1in}\\
\epsfig{file=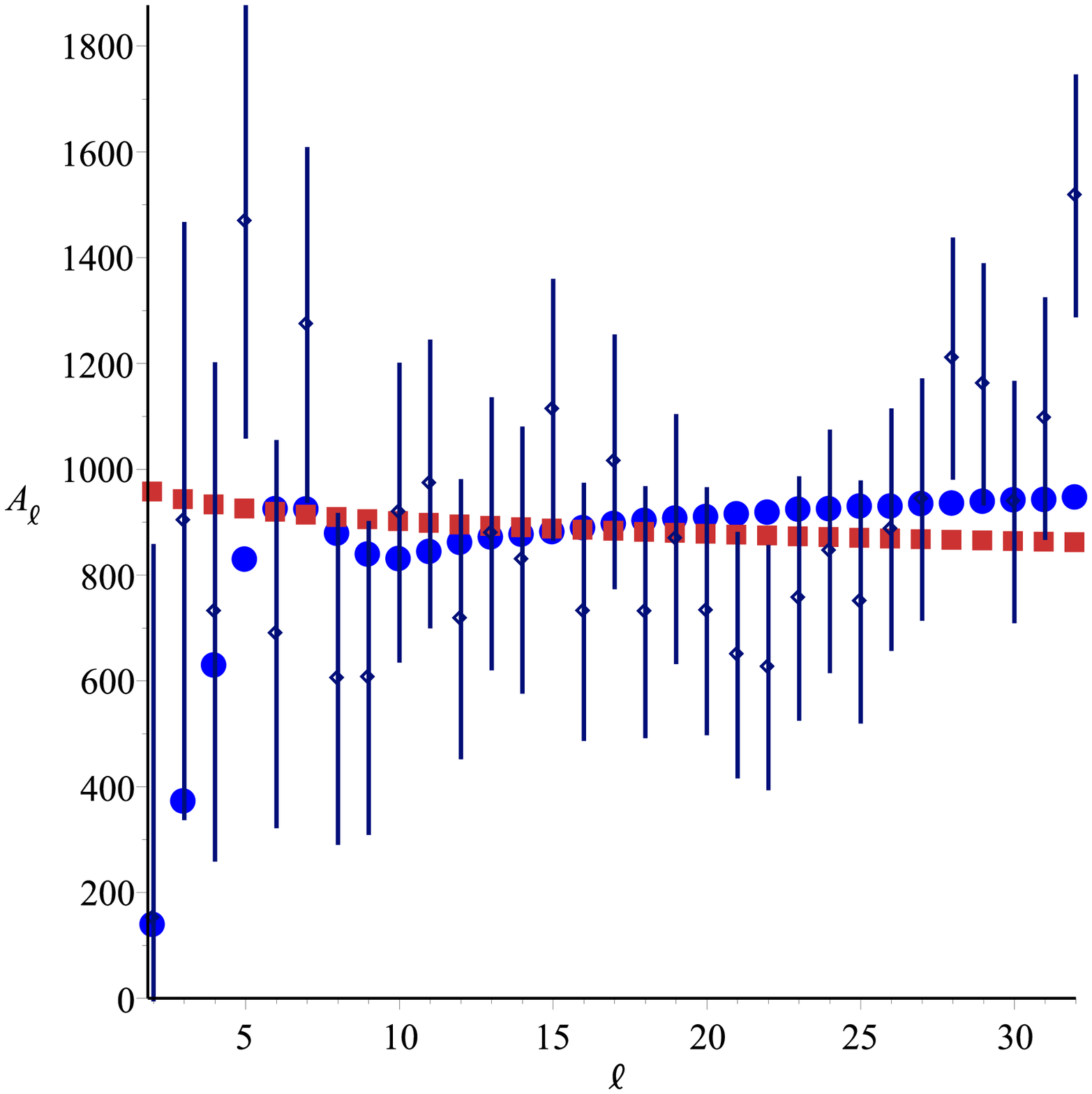, height=1.1in, width=1.1in}& \qquad\quad
\epsfig{file=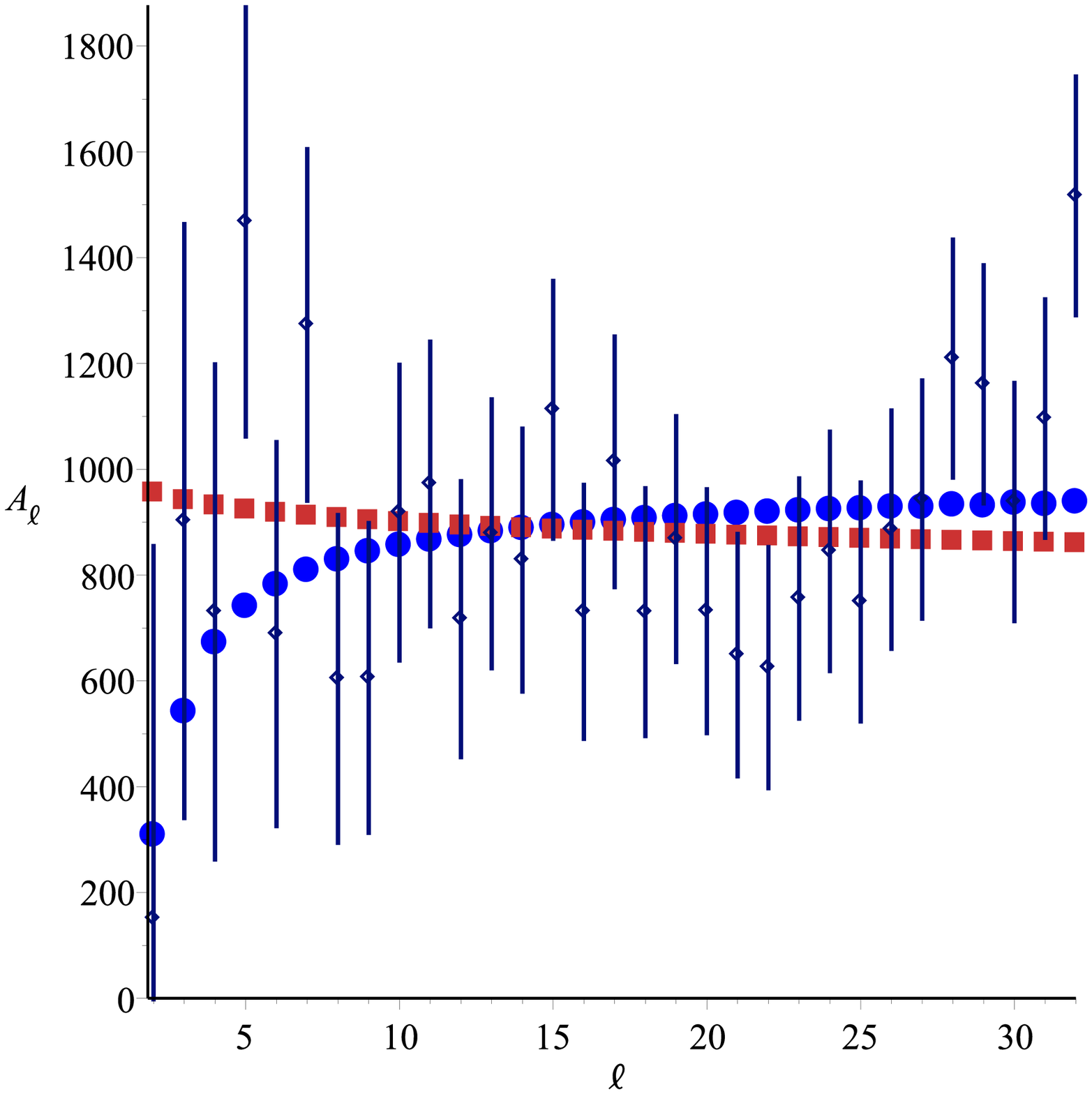, height=1.1in, width=1.1in}& \qquad\quad
\epsfig{file=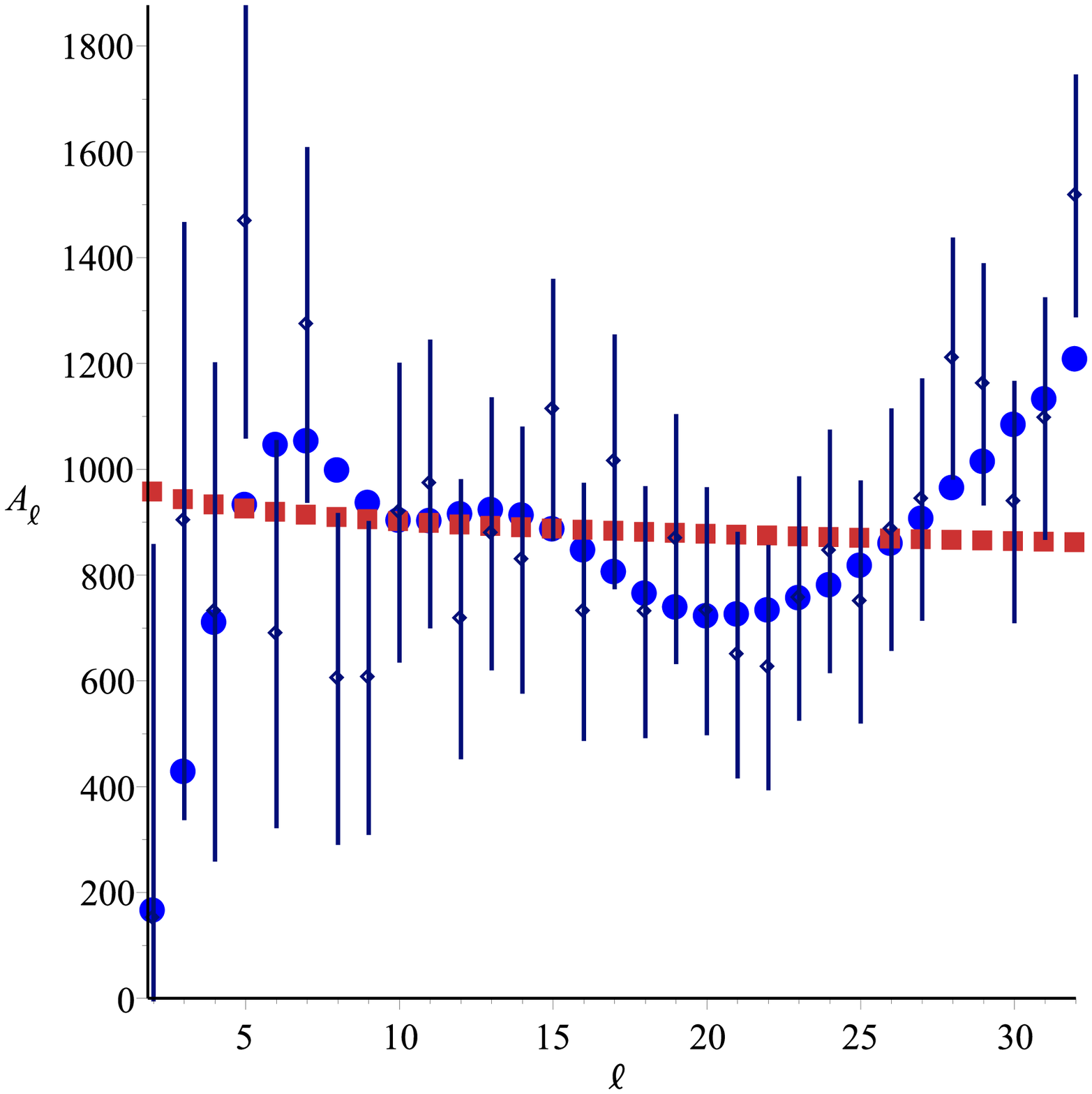, height=1.1in, width=1.1in}
\end{array}$
\end{center}
\caption{\small
The dependence of $\chi^{\,2}(\varphi_0,{\cal M},\delta)$, minimized with respect to $M$, on $\delta$, for three significant cases corresponding to the first three collections of power spectra in fig.~\ref{fig:double} (first row), and the corresponding optimal angular power spectra (second row).}
\label{fig:chi2delta}
\end{figure}

In order to compare these power spectra with the low--$\ell$ CMB, in \cite{ks} we followed a simple procedure guided by the order parameter
\beq
\chi^2(\varphi_0,{\cal M},\delta) \ = \ \ \sum_{\ell=2}^{32} \frac{\left(A_\ell(\varphi_0,{\cal M},\delta)\ - \ A_\ell^{\rm WMAP9}\right)^2}{\left(\Delta A_\ell^{\rm WMAP9}\right)^2} \ ,
\eeq{33}
where the ${\cal A}_\ell$ were computed resorting to eq.~\eqref{4} of the Introduction \cite{cmb_slow},
\beq
A_\ell\;(\varphi_0,{\cal M},\delta) \ = \ {\cal M} \ \ell(\ell+1)\ \int_0^\infty \frac{dk}{k} \ {\cal P}_R \big( k , \varphi_0 \big) \, {j_\ell}^2 \big( k \, 10^\delta \big) \ ,
\eeq{34}
in terms of the normalization ${\cal M}$ and of a second parameter, $\delta$. ${\cal M}$ fixes essentially the energy scale of the process, and was determined, in all cases that we examined, minimizing analytically $\chi^2(\varphi_0,{\cal M},\delta)$. The second parameter is more interesting: it effects an horizontal displacement, in semi--logarithmic scale, of the power spectra of fig.~\ref{fig:double}, making it possible to optimize their correlation with features present in the angular power spectra of fig.~\ref{fig:compared_CMB}.
\begin{figure}[ht]
\begin{center}$
\begin{array}{ccc}
\epsfig{file=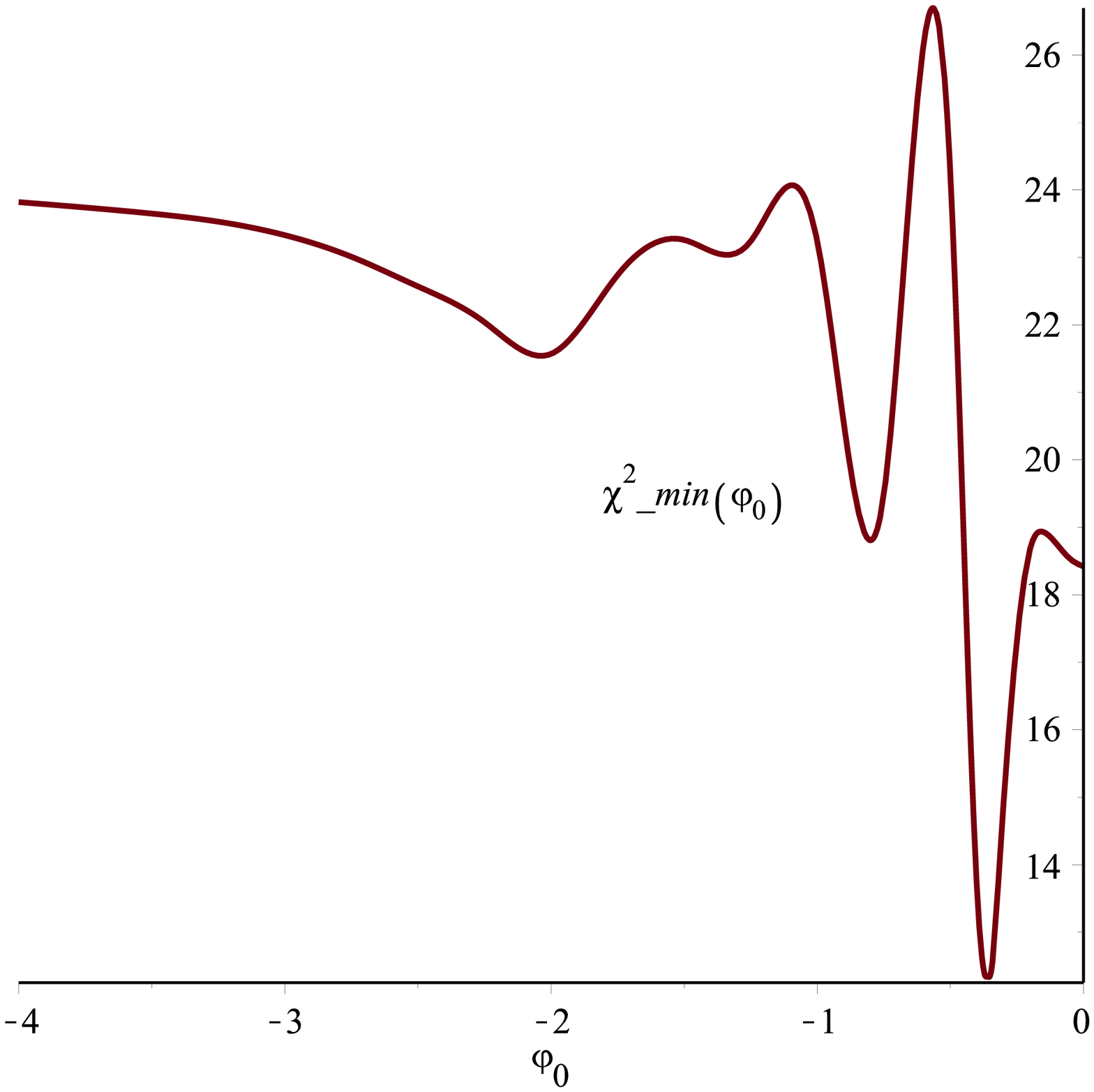, height=1in, width=1.1in}& \quad\qquad
\epsfig{file=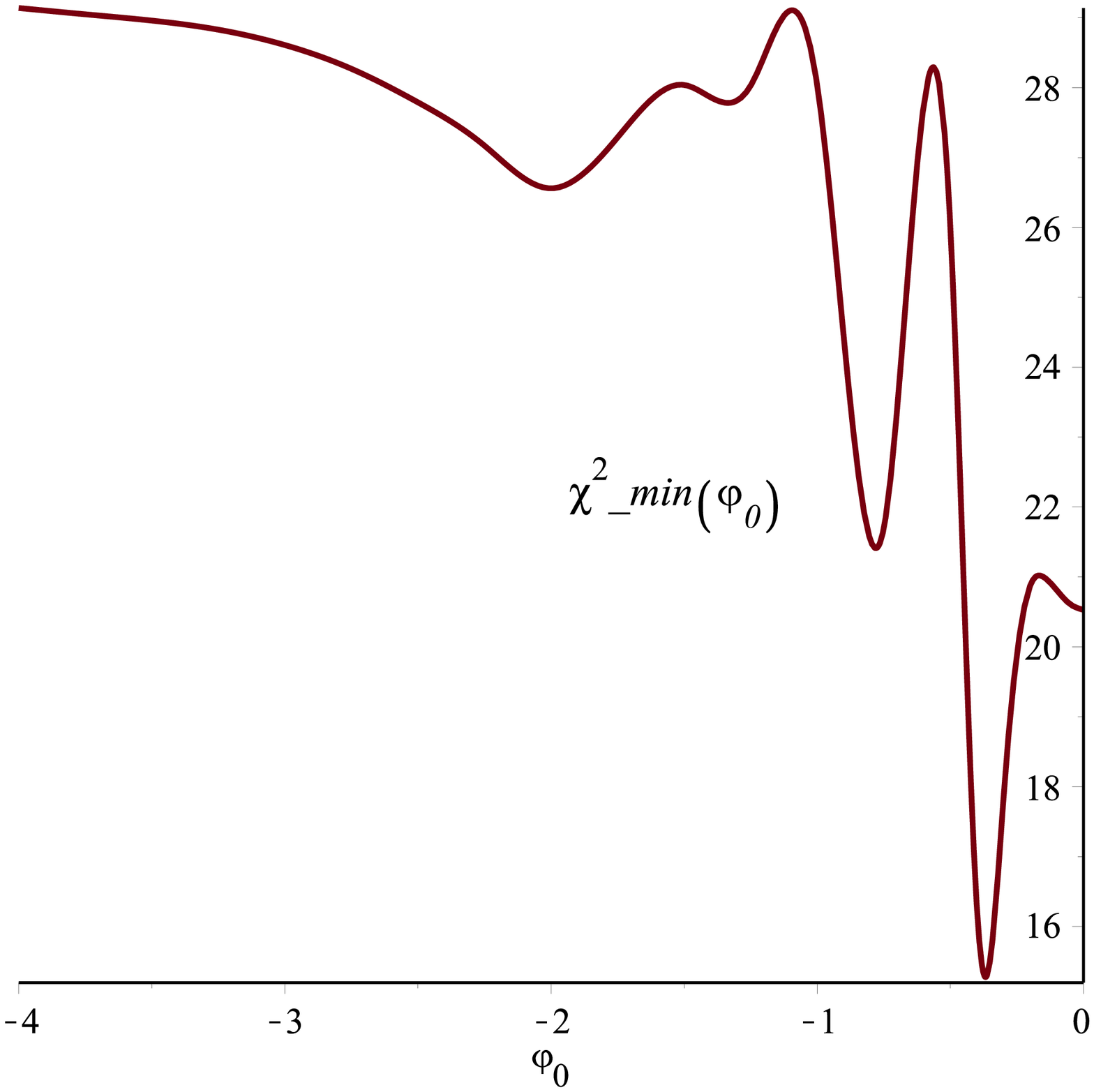, height=1in, width=1.1in}& \quad\qquad
\epsfig{file=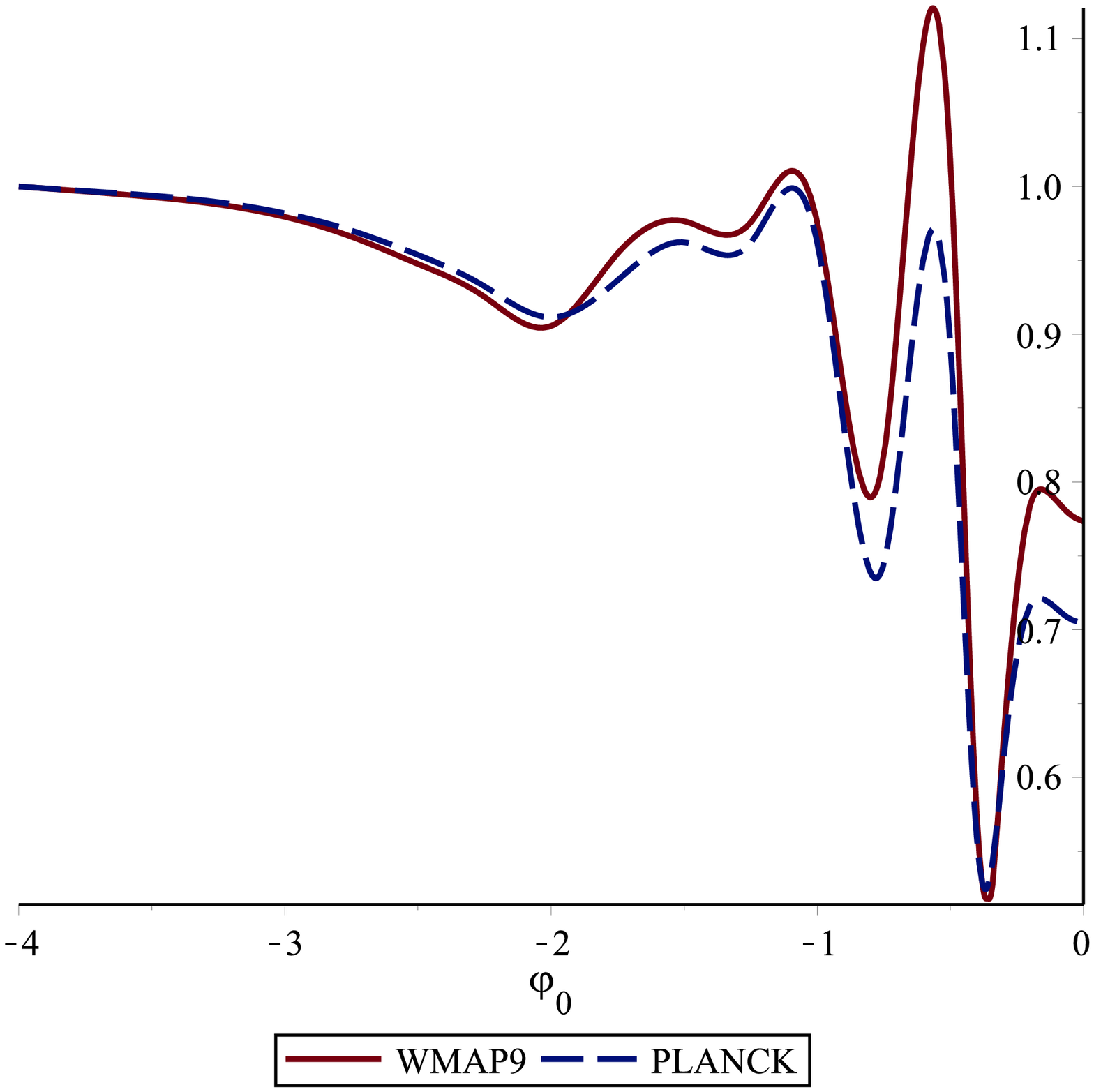, height=1in, width=1.1in}
\end{array}$
\end{center}
\caption{\small
The optimal values of the minimum values of $\chi^2(\varphi_0,{\cal M},\delta)$ as a function of $\varphi_0$ for WMAP9 data (left), PLANCK 2013 data (center) and their comparison (right) when both are normalized with respect to their values for $\varphi_0=-4$, which are respectively 23.82 and 29.14.}
\label{fig:angular_fit_double_bump}
\end{figure}

The simple order parameter $\delta$ displays interesting behaviors, and the reader will not fail to recognize amusing analogies with Landau--Ginzburg theory. The left portion of fig.~\ref{fig:chi2delta} corresponds to an intermediate case in the left portion of fig.~\ref{fig:double}, where the pre--inflationary peak has barely separated from the almost scale invariant profile. The shallow but narrow minimum met when varying $\delta$ identifies a horizontal displacement that hooks the pre--inflationary peak to the features present around $\ell=5$. The central portion corresponds to the central portion of fig.~\ref{fig:double}, and more precisely to the featureless spectrum displayed there. Here the shallow and wide minimum met when varying $\delta$ is driven by the quadrupole depression. Finally, the right portion corresponds to the right portion of fig.~\ref{fig:double}, and now a deep, narrow minimum signals a stronger correlation between the features around $\ell=5$ and the dip around $\ell=20$. The results obtained with WMAP9 and {\sc Planck} 2013 are nicely consistent, up to an overall normalization, as can be seen in fig.~\ref{fig:angular_fit_double_bump}, whose entries display the dependence on $\varphi_0$ of the minimum value attained by $\chi^2(\varphi_0,{\cal M}, \delta)$ as one varies ${\cal M}$ and $\delta$. Although the ratio $\chi^2/DOF$ between this minimum $\chi^2$ and the actual number of unconstrained parameters can drop below 50\% of its value for eq.~\eqref{1}, as one can see from fig.~\ref{fig:double} these models suffer from a slow approach to the power spectrum of eq.~\eqref{1} that makes them, at most, interesting toy examples. On the other hand, confining the attention to quadrupole depression and paying some attention to the Galactic mask, as I can now explain, one can come closer to a more realistic, if still partial, description of low--$\ell$ anomalies.

\scs{Full spectrum analysis and the Galactic mask} \label{sec:galactic}
%
\begin{figure}[ht]
\centering
\begin{tabular}{c}
\includegraphics[width=80mm]{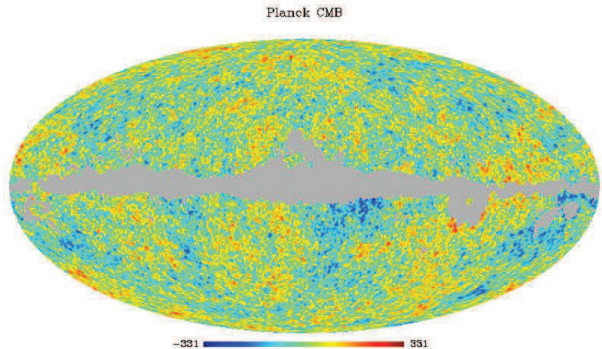}
\end{tabular}
\caption{\small All studies of the CMB rest on a choice of Galactic mask, aimed at reducing contaminations while granting access to the largest possible fraction of the sky. The mask used in {\sc Planck} 2015 corresponds to $f_{sky}=94\%$, \emph{i.e.} it grants access to 94\% of the sky.}
\label{fig:galactic}
\end{figure}

In this section, I would like to summarize some recent steps leading to an extension of $\Lambda$CDM where a new parameter, a scale $\Delta$, accounts for the apparent lack of power at low $\ell$. Adding one more parameter in full-fledged likelihood tests is a delicate operation, which could easily destabilize the celebrated determination of the others or lead to inconclusive results.
Eqs.~\eqref{25} suggest a convenient way to proceed, since \emph{$W_S$ must cross the $\eta$--axis as a system that started out in a kinetically dominated phase approaches slow roll}.

A convenient way to model the effect rests on the MS equation for the Coulomb--like potentials
\beq
W_S = \frac{\nu^2-\frac{1}{4}}{\eta^2}\left[c \left(1+\frac{\eta}{\eta_0}\right)\ + \ (1-c) \left(1+\frac{\eta}{\eta_0}\right)^2\right] \ ,
\eeq{35}
which approach for $\eta \to 0^-$ the attractor behavior of eq.~\eqref{28} with
\beq
n_s \ = \ 4 \ - \ 2\, \nu \ ,
\eeq{361}
and change sign at $\eta=-\eta_0$ with a slope determined by $c-2$. The corresponding power spectra can be computed exactly \cite{dkps}:
\beq
{\cal P}_{R} (k) \ \sim \ \frac{{(k\,\eta_0)^3}\, \exp\left(\frac{\pi \left(\frac{c}{2}\, -\, 1\right)\left(\nu^2 \, - \, \frac{1}{4}\right)}{\sqrt{\left(k \,\eta_0\right)^2 \ + \ (c\, -\, 1)\left(\nu^2 \, -\,  \frac{1}{4}\right)}} \right) }{\left|\Gamma\left(\nu \, + \, \frac{1}{2} \, + \, \frac{i\,\left(\frac{c}{2}\, -\, 1\right)\left(\nu^2 \, - \, \frac{1}{4}\right)}{\sqrt{\left(k \,\eta_0\right)^2 \ + \ (c\, -\, 1)\left(\nu^2 \, - \, \frac{1}{4}\right)}}\right)\right|^2\, { \left[\left(k \,\eta_0\right)^2 \ + \ (c\, -\, 1)\left(\nu^2 \, - \, \frac{1}{4}\right)\right]^{\nu}}} \ .
\eeq{36}
In particular, for $c=2$ this expression reduces to a simple modification of eq.~\eqref{1}, which I would like to present in the form
\beq
{\cal P}_R(k) \ = \ A \ \left(k/k_0\right)^{n_s-1} \ \rightarrow \ \frac{ A \, \left(k/k_0\right)^3}{\left[\left(k/k_0\right)^2 + \left(\Delta/k_0\right)^2\right]^{\nu}} \ ,
\eeq{37}
where $k_0$ is a pivot scale. For $c>2$ the power spectra in eq.~\eqref{36} embody pre--inflationary peaks reminiscent of those in the first and fourth groups of plots of fig.~\ref{fig:double}.
\begin{figure}[ht]
\centering
\begin{tabular}{cc}
\includegraphics[width=45mm]{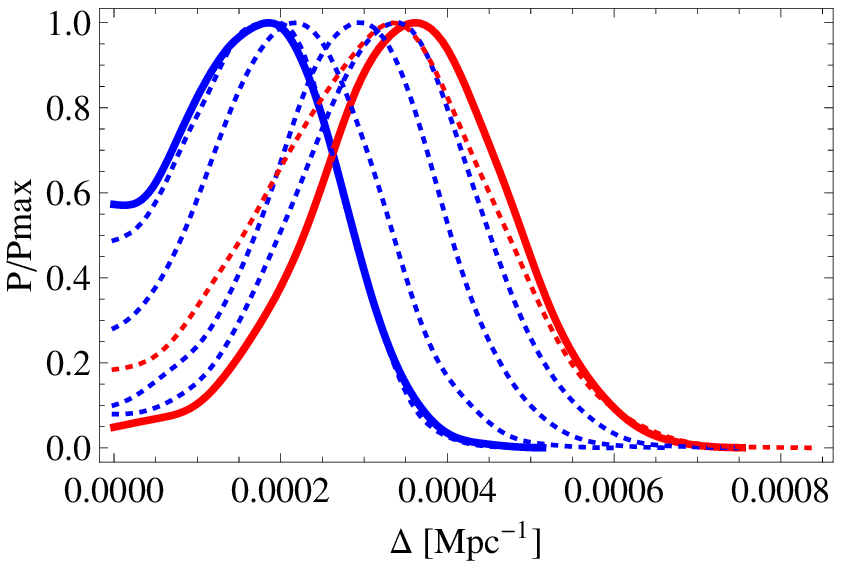} \qquad\qquad &
\includegraphics[width=50mm]{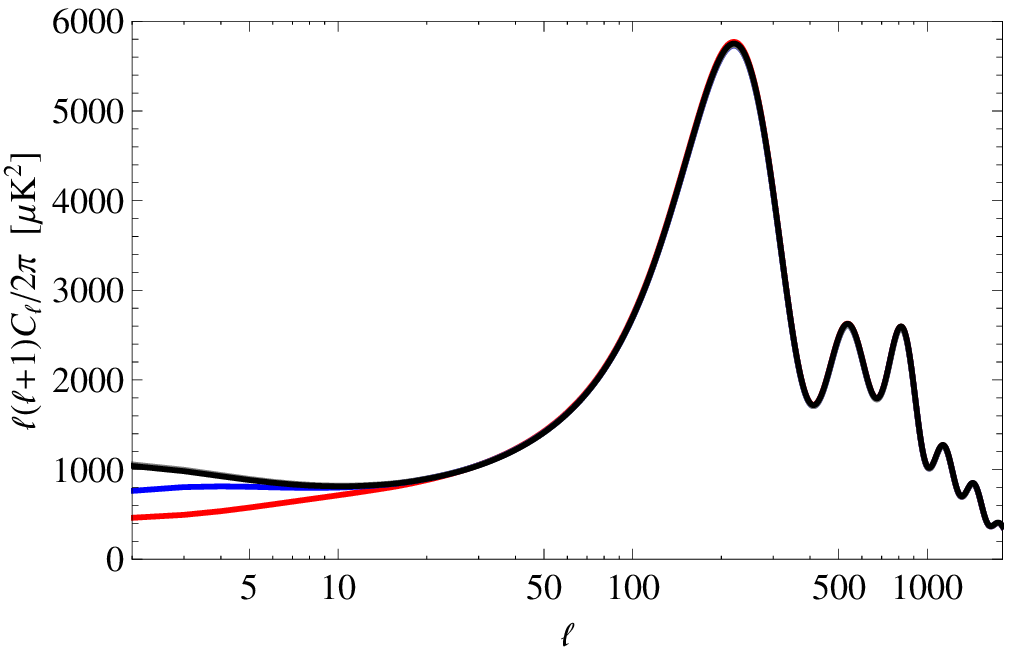}
\end{tabular}
\caption{\small {\it Left:} posteriors for the parameter $\Delta$ in the power law (\ref{37}), for different Galactic mask: solid blue for the 94\% mask, thick red for a $+30^\circ$ extension, dotted blue for the intermediate masks $+6^\circ$, $+12^\circ$, $+18^\circ$, $+24^\circ$, and dotted red for $+36^\circ$.
{\it Right}: best fit APS models for $\Lambda$CDM with a standard mask (black), for $\Lambda$CDM with a $+30^\circ$ extension (gray, barely visible behind the previous case),
for $\Lambda$CDM$+\Delta$ with a standard mask (blue), and for $\Lambda$CDM$+\Delta$ with a $+30^\circ$ extension (red).}
\label{fig:compared_Delta}
\end{figure}

I can now summarize a few relevant facts \cite{gs}:
\begin{itemize}
\item  the detection of $\Delta$ improves with wider masks obtained \emph{extending blindly} the standard {\sc Planck} mask that leaves out a fraction $f_{sky}=94\%$, up to $f_{sky}=39\%$, and then starts to deteriorate. At the same time, its mean value nearly doubles. If only $\Delta$ is introduced, the determination of the standard cosmological and nuisance parameters is not sensibly affected, even if polarization data are taken into account. On the other hand, if $\Delta$ is not introduced, widening the mask does affect the determinations to some extent. In \cite{gs} we thus obtained, with $f_{sky}=39\%$, the optimal detection
\beq
\Delta = (0.351 \pm 0.114) \times 10^{-3} \, \mbox{Mpc}^{-1} \ ,
\eeq{38}
which corresponds to about 99\% CL and to the red curves in fig.~\ref{fig:compared_Delta}. As we stressed in \cite{gs}, this result should be compared to a nearby value obtained, at 88\% CL, in the standard mask with $f_{sky}=94\%$;

\item with also $c$ in eq.~\eqref{36}, the joint determination of both parameters becomes less effective, while the preferred $c$ is compatible with a mild pre--inflationary peak;

\item the idea that a short inflation might have left some imprints in the CMB APS was already pursued, in interesting directions, in a number of papers including those listed in \cite{standard_peak}, and more recently it was also considered in \cite{standard_peak_recent}. We came to it from a different perspective in \cite{dks,dkps,ks}, and the work in \cite{gs} differs from previous or other present approaches in three respects: the well--motivated parametrization in eqs.~\eqref{36} and \eqref{37}, the use of the best available data from {\sc Planck} 2013 and  from {\sc Planck} 2015, including polarization effects, and above all the use of a sequence of wider Galactic masks. The end result is the relatively good evidence for the scale $\Delta$ in eq.~\eqref{38}. Clearly, the significance of the wider masks deserves further investigations.
\end{itemize}

\begin{figure}[ht]
\centering
\begin{tabular}{cc}
\includegraphics[width=35mm]{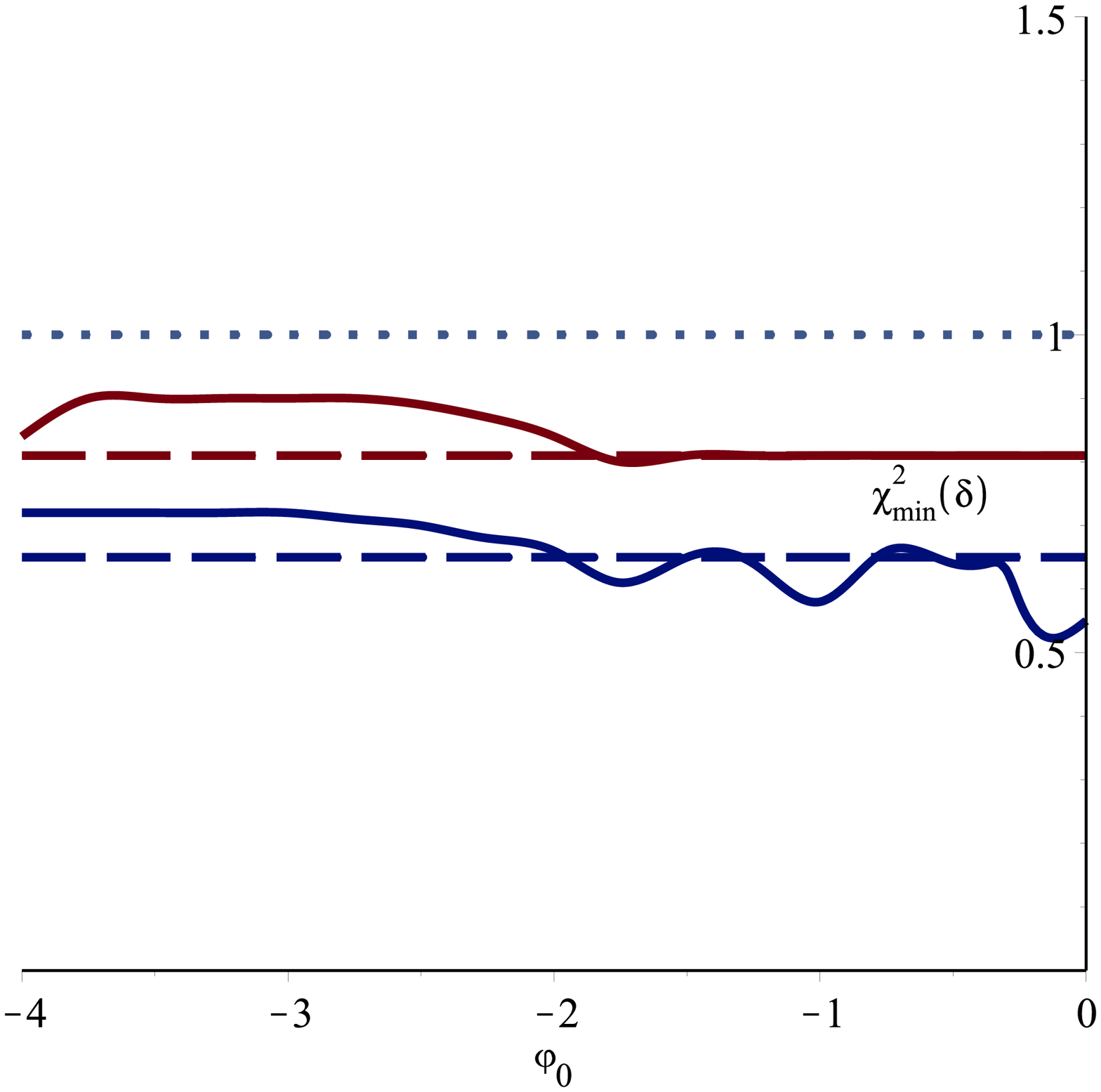} \qquad\qquad  &
\includegraphics[width=35mm]{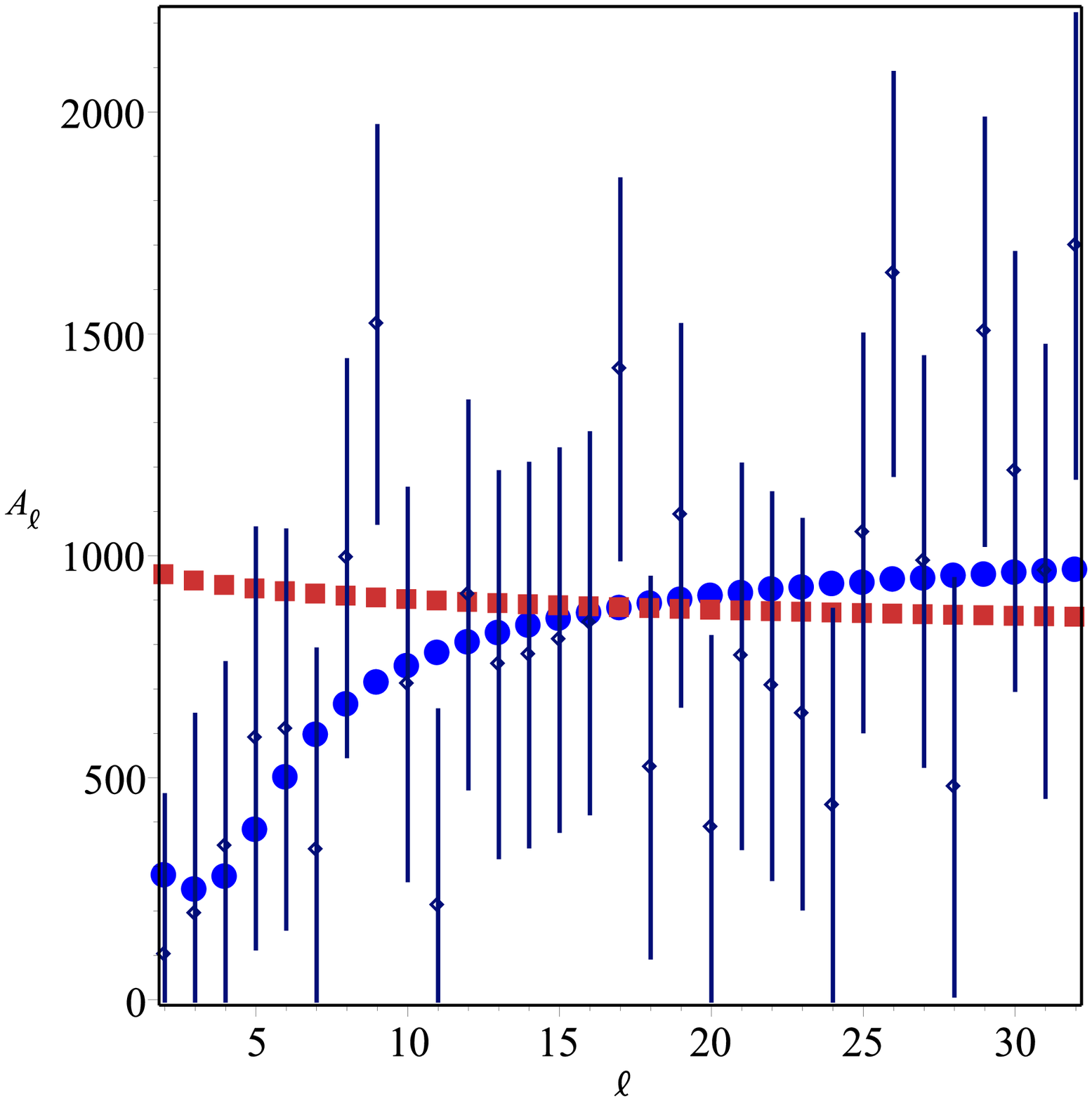}\vspace{12pt}
\end{tabular}
\caption{\small {\it Left:} the dependence on $\varphi_0$ of $\chi^2(\varphi_0,{\cal M}, \delta)$ minimized in ${\cal M}$ and $\delta$ fades out with {\sc Planck} 2013 data (red, continuous) or {\sc Planck} 2015 data (blue, continuous) and a $+30^\circ$ mask, corresponding to $f_{sky}=39\%$. {\it Right:} the {\sc Planck} 2015 low--$\ell$ APS and the optimal spectrum in the third group of fig.~\ref{fig:chi2delta}, which becomes nearly indistinguishable from that determined by eq.~\eqref{37}.}
\label{fig:masked_chi2}
\end{figure}

I would like to conclude with some unpublished material that I presented in preliminary form at Erice and more fully at Sestri Levante. It is aimed at justifying, within a simple $\chi^2$ analysis as in Section \ref{sec:chi2}, the peculiar effect of the Galactic mask on the determination of $\Delta$. Referring to fig.~\ref{fig:masked_chi2}, one should keep in mind that the primordial power spectra of fig.~\ref{fig:double} are in semi--logarithmic scale. Widening the masks cuts out oscillations in the first multipoles, as can be seen comparing figs.~\ref{fig:compared_CMB} for {\sc Planck} 2015 and \ref{fig:masked_chi2}, leaving behind a slower growth. This increases $\Delta$, shifting to the right ${\cal P}_R(k)$, in semi-logarithmic scale, via the parameter $\delta$. However, this also \emph{dilates} its profile in the linear scale of eq.~\eqref{4}, washing out pre--inflationary peaks. As a result, very different power spectra yield APS that are essentially indistinguishable, for $\ell \leq 32$, from those resting on the analytic profile in eq.~\eqref{37}.

\subsection*{Acknowledgements} I would like to thank Prof.~A.~Zichichi for kindly inviting me to lecture at Erice. I am also grateful to the CPhT--E.~Polytechnique and to the CERN Ph-Th Department for their kind hospitality during my sabbatical year, and to E.~Dudas, A. Gruppuso, A.~Kehagias, N. Kitazawa, N.~Mandolesi, P.~Natoli, M.~Porrati, R.~Stora, A.~Yeranyan, and especially to S.~Ferrara, for very stimulating collaborations.
This work was supported in part by Scuola Normale Superiore and by INFN-CSN4 (IS Stefi).

\end{document}